\pdfoutput=1
\documentclass[aps,prb,twocolumn,longbibliography,superscriptaddress]{revtex4-2}
\usepackage{amsmath}

\usepackage{bm}
\usepackage{bbold}
\usepackage{amssymb}
\usepackage{graphicx}
\usepackage[ hidelinks = true, colorlinks = true, citecolor = blue, urlcolor = blue, linkcolor = blue]{hyperref}
\usepackage{verbatim}
\usepackage{xcolor}
\usepackage[capitalise]{cleveref}
\usepackage[normalem]{ulem}
\usepackage{dsfont}

\usepackage{accents}

\newcommand{\ket}[1]{\left|  #1 \right>}
\newcommand{\bra}[1]{\left<  #1 \right|}

\begin{document}

\title{ Dispersive detection of single microwave photons  with quantum dots}

\author{Stephanie Matern}
\affiliation{CNR -- INO Pitaevskii BEC Center and Dipartimento di Fisica, Universit\`a di Trento, Trento, Italy}
\author{Alberto Biella}
\affiliation{CNR -- INO Pitaevskii BEC Center and Dipartimento di Fisica, Universit\`a di Trento, Trento, Italy}
\affiliation{INFN-TIFPA, Trento Institute for Fundamental Physics and Applications, Via Sommarive 14, I-38123 Trento, Italy}
\author{Pasquale Scarlino}
\affiliation{Institute of Physics and Center for Quantum Science and Engineering,Ecole Polytechnique F{\'e}d{\'e}rale de Lausanne, CH-1015 Lausanne, Switzerland}
\author{Iacopo Carusotto}
\affiliation{CNR -- INO Pitaevskii BEC Center and Dipartimento di Fisica, Universit\`a di Trento, Trento, Italy}
\affiliation{INFN-TIFPA, Trento Institute for Fundamental Physics and Applications, Via Sommarive 14, I-38123 Trento, Italy}
\author{Gianluca Rastelli}
\affiliation{CNR -- INO Pitaevskii BEC Center and Dipartimento di Fisica, Universit\`a di Trento, Trento, Italy}
\affiliation{INFN-TIFPA, Trento Institute for Fundamental Physics and Applications, Via Sommarive 14, I-38123 Trento, Italy}

\date{\today}

\begin{abstract}
Within a  circuit quantum electrodynamics architecture, we theoretically investigate the detection of a single propagating microwave photon traveling through a resonant microwave cavity dispersively interacting with a double quantum dot tunnel-coupled to a lead.
Under suitable conditions, a single photon in the cavity can induce a measurable change in the electronic occupation of the charge states.
We develop a quantum cascade approach that enables a time-resolved description of a single-photon wave packet impinging on the cavity.
We make use of a simple model of charge detector to assess the efficiency of our photo-detection configuration as functions of key parameters such as coupling strength, tunneling rate, temperature, and photon resonance linewidth.
We finally highlight a measurement-induced backaction effect on the cavity mode associated with the dispersive, non-absorptive detection process.
\end{abstract}

\maketitle

\section{Introduction}
Circuit quantum electrodynamics (circuit QED) \cite{Haroche:2020,Blais:2021,Gu:2017} has progressed through remarkable advances in superconducting devices, which not only serve as qubits for quantum information processing but can also be used for manipulation of quantum microwave fields. 
Despite major advances in microwave engineering,  quantum photonics in this spectral domain remains at a much earlier stage compared to its optical counterpart. 
For instance, detecting a single itinerant microwave photon remains challenging due to its low energy.
This represents a critical gap, as single-photon detection is fundamental in quantum optics \cite{Hadfield:2009}, playing a key role in correlation measurements \cite{mandel1995optical,walls2025quantum}, 
linear optics quantum computing \cite{RevModPhys.79.135}, and remote entanglement protocols \cite{hermans2023entangling}.

Using Josephson junction circuits, single microwave-photon detectors based on absorption processes have been realized through various schemes. 
One approach exploits metastability, as in current-biased Josephson junctions \cite{Chen:2011,Oelsner:2013,Oelsner:2017,Kuzmin:2018,Revin:2020,Pankratov:2022,DElia:2023} acting as switching detectors \cite{Poudel:2012,Andersen:2013,Andersen:2014,Golubev:2021}, in SQUID devices with a double-well potential \cite{Opremcak:2018,Opremcak:2021}, or in Kerr Josephson parametric resonators \cite{Petrovnin:2024}.  
Another approach relies on engineered dissipation, such as impedance-matched $\Lambda$ systems in 1D waveguide circuit QED \cite{Koshino:2009,Koshino:2013,Koshino:2015,Inomata:2016}, 
inelastic tunneling of quasiparticles 
in voltage-biased 
superconducting  junctions  
\cite{Stanisavljevi:2024,Basset2025}, or four-wave mixing systems incorporating a strongly damped cavity mode \cite{Lescanne:2020,Albertinale:2021,Wang:2023,Billaud:2022,Braggio:2025}.
Moreover, bolometers scheme have bee also applied \cite{Govenius:2016,Kokkoniemi:2020,Lee:2020}. 
Additionally, non-absorbing detection has been demonstrated in pulsed-mode operation using a controlled-phase gate between a superconducting qubit and a propagating photon \cite{Kono:2018,Besse:2018}, or by reaching the strong dispersive regime  \cite{Narla:2016}.

New perspectives for single microwave photons detection  
are offered by the combination of circuit QED systems with quantum dots (QDs). These act as artificial atoms \cite{Chatterjee:2021,Zhang:2019,Burkard:2023} 
interacting with microwave photons in superconducting circuits \cite{Cottet:2017review,Burkard:2020}.  
The uniqueness of these systems lies in the seamless integration of quantum-coherent nanoscale devices, where semiconducting quantum nanostructures can be combined with superconducting elements and interfaced with a microwave photonic system \cite{Frey:2011,Delbecq:2011,Petersson:2012,Basset:2013,Braakman:2014,Rossler:2015,Viennot:2015,Ranjan:2015,Viennot:2016,Beaudoin:2016,Bruhat:2016}. 
They provide a framework to study light–matter interactions in the regime of coherent, far-from-equilibrium single-electron transport 
\cite{Brandes:2003gu,Childress:2004kt,Trif:2008,Cottet_Kontos:2010,Cottet_Kontos_Yeyati:2012,Jin:2012,Hu:2012,Bergenfeldt:2013,Contreras-Pulido:2013,Kloeffel:2013,Bergenfeldt:2014,Cottet:2015a,Dmytruk:2016,Cottet:2020,Governale:2025,Hellbach:2022,Rastelli:2019,Mantovani:2019a,Mantovani:2019b}.
In particular, the regime of strong light-matter coupling has been reached for both charge  \cite{Mi:2016,Stockklauser:2017,Mi:2017,Scarlino:2019,Li_Li_Gao:2018} 
and spin qubits \cite {Samkharadze:2018,Mi:2018ip,Landig:2018,Zihlmann:2023}.

Early experiments showed that  the absorption of a single photon can induce 
photo-assisted single-electron tunneling 
events \cite{Kouwenhoven:1994,Oosterkamp:1998,Gustavsson:2007,Leturcq:2008}.
The basic photon detection scheme has been studied in a double quantum dot (DQD) resonantly coupled to a microwave cavity mode \cite{Wong:2017, Ghirri:2020}: 
Photon absorption promotes an electron from the DQD ground (bonding) state to an excited (antibonding) state, from which it can tunnel to the leads.
In this way, for asymmetric tunneling between the DQD and the left and right contacts, a steady photon flux generates a continuous, detectable charge current through the DQD.
The experimental realization of this basic scheme was reported in \cite{Khan:2021} and in following related experiments \cite{Cornia:2023,Haldar:2024,Haldar:2024b, Haldar:2024c}: While a proof of principle of the process was demonstrated, the detection efficiency $\eta$, defined as the ratio of photocurrent to photon flux, remained modest.
A far improved efficiency $\eta \sim 70\%$ was reported in a later experiment~\cite{Oppliger:2025}. 

All these experiments were based on photocurrent detection through a DQD, but individual photon-absorption events remain  challenging to detect using current-based schemes. 
Single-electron tunneling events typically generate very short current pulses with amplitudes several orders of magnitude below the noise floor of available electronic instrumentation \cite{Elzerman:2003}.
To overcome this limitation and enable detection of single-photon absorption events, an alternative approach was 
explored in \cite{Haldar:2024a}, where charge-state switching in  DQDs, triggered by a single absorbed photon, is monitored in real time using a highly sensitive, fast charge sensor capacitively coupled to the dots. 
Although promising as strategy, the efficiency reported in this work remained as small as a few percent.
Achieving a continuous mode single-photon microwave detection  continues therefore to represent a significant open challenge in the field.

\begin{figure}[t]
\centering
\includegraphics[width=0.95\linewidth]{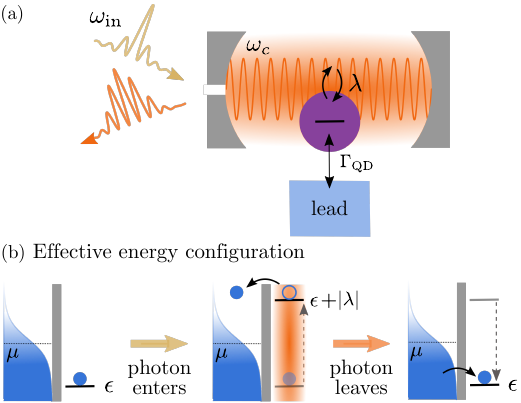}
\caption{
(a) 
The system consists of an incoming microwave photon pulse with frequency $\omega_\textrm{in}$, which is temporarily stored in a cavity (detection mode with frequency $\omega_c$). 
The cavity is dispersively coupled (coupling strength $\lambda$) to a single electron state, 
from which an electron can tunnel to and from a single lead at temperature $T$ and chemical potential $\mu$ with rate $\Gamma_\textrm{QD}$.
(b) 
Concept of the dispersive measurement by monitoring the electron occupation. 
Initially, the energy of the electronic state is $\epsilon < \mu$, so this is almost fully occupied (in the low-temperature limit, see main text). 
When a photon enters the cavity, the energy of the electronic state shifts $\epsilon \to \epsilon + |\lambda|$, and the electron 
can tunnel out into the lead. 
The system charge occupation is monitored through a charge detector 
(not shown in the figure).
Here we assumed the dispersive coupling $\lambda<0$ (see main text).
The single electron level corresponds to the bonding level of the DQD whereas the antibonding level (not shown) always remains empty,
see \cref{fig:app_DQD_levels} in Appendix \ref{app:DispersiveRegime}.
}   \label{fig:system}
\end{figure}

In this work, we analyze a minimal setup for achieving dispersive single-microwave-photon detection. 
The scheme leverages
the experimental feasibility of monitoring the charge states  in real time.
The basic idea is illustrated in Fig.~\ref{fig:system}, providing a qualitative picture as described below.
The effective model reduces to a single-electron level laterally tunnel-coupled to a lead (corresponding to the bonding charge state of the DQD, see  \cref{app:DispersiveRegime}). 
Assuming an initial state in which the electronic level is well below the lead's chemical potential, the electronic state is fully occupied.
When the photon enters the cavity, it induces  
a positive energy shift of the electronic state. 
If this shift is large enough, the electronic level moves above the chemical potential, allowing the electron to tunnel out to the lead.
Once the photon leaves the cavity, the electronic state returns to its initial energy value, and it gets refilled by an electron tunneling in from the lead. 
Although this simplified picture strictly holds only at vanishing temperature, it motivates a detailed analysis of the system dynamics at finite temperature, to explore the parameter ranges (temperature, interaction strength, etc.) and assess its actual promise as photon detector.

%%%%%%%%%%%%%%%%%%%%%%%%%%%%%%%%%%%%%%%%%%%%%%%%%%%%%%%%%%%%%%
%
%
%   FIGURE N. 1 
%
%
%
%
%
%%%%%%%%%%%%%%%%%%%%%%%%%%%%%%%%%%%%%%%%%%%%%%%%%%%%%%%%%%%%%%%

The paper is structured as follows. 
In \cref{sec:model} we describe the effective model, formed 
by a single electron level tunnel-coupled to a macroscopic lead and dispersively interacting with a resonant mode of a microwave cavity. 
The full quantum dynamics of the system is governed by a Lindblad-like master equation.
To model the dynamics of  a single-photon wavepacket traveling towards the cavity, 
we use the quantum cascade approach.
For the electron tunneling to the lead, 
we derive the corresponding 
dissipative superoperator in presence 
of the dispersive interaction with the microwave mode. 
In \cref{sec:QND}, we present the results for the quantum dynamics of the system in the single-photon regime, 
explaining the behavior of the cavity and electron occupations as functions of the relevant parameters and highlighting a measurement-induced backaction effect on the cavity photon occupation.
In \cref{sec:efficiency}, 
we analyze the detection efficiency based on 
a simple phenomenological model for the charge detector.
In \cref{sec:discussions}, we critically discuss the challenges and experimental feasibility of our proposal.
Finally, we draw our conclusions in \cref{sec:conclusions}.

%%%%%%%%%%%%%%%%%%%%%%%%%%%%%%%%%%%%%%%%%%%%%%%%%%%%%%%%%%%%%%%%%%%%%%%%
%
%
%   SECTION 1 : THEORETICAL MODEL
%
%

\section{Theoretical model} 
\label{sec:model}
The dispersive interaction between the DQD and the cavity mode 
reads 
\begin{equation}
\hat{H}_\textrm{dis} = 
  \omega_c \hat{n}_c 
+
\epsilon  \, \hat{n}_\textrm{QD}
+
\epsilon'  \, \hat{n}_\textrm{QD}'
+
\lambda \, \hat{n}_c \left( \hat{n}_\textrm{QD}'-  \hat{n}_\textrm{QD}  \right),
\label{eq:main_Hdis}
\end{equation}
with the energy $\epsilon,\epsilon'$ of the bonding and the antibonding levels respectively,  $\omega_c$ is the frequency of the cavity mode and dispersive interaction strength $\lambda$.
Hereafter we denote the resonant mode coupled to the DQD as the detection mode.  
The number operators read 
$\hat{n}_\textrm{QD} = \hat{d}^{\dagger} \hat{d}^{\phantom{g}} $
($\hat{n}_\textrm{QD}'= \hat{d}'^{\dagger} \hat{d}'$)
and 
$\hat{n}_c =\hat{b}^{\dagger} \hat{b}^{\phantom{g}} $  
%            and 
%           $\hat{n}_{\textrm{in}} = \hat{a}^{\dagger} 
%           \hat{a}^{\phantom{g}} $  
%
in terms of the creation and annihilation operators 
of the  electron levels (fermionic operators) and the detection mode (bosonic operators). 
We set $\hbar = k_B = 1$ throughout the paper.

A derivation of the effective dispersive Hamiltonian of \cref{eq:main_Hdis} 
from the full Rabi model, a DQD coupled to a single microwave mode, is given in \cref{app:DispersiveRegime}
(see also Refs.~\cite{Zueco:2009,Zhu:2013,Clemens:2020}).
There, we also show that in a suitable parameter regime 
the antibonding state is irrelevant as its occupation remains unaffected by the photon occupation in the cavity (see \cref{fig:app_DQD_levels}).
Then we can study 
the effective system composed by a single electronic level
(the bonding state of the DQD) whose energy depends on the cavity photon occupation described by the following Hamiltonian 
%
%%%%%%%%%%%%%%%%%%%%%%%%%%%%%%%%%%%%
%
%   EQUATION N. 2
%
%
\begin{equation}
\label{eq:H_s}
\hat{H}_S = 
  \omega_c \hat{n}_c 
+
\epsilon  \, \hat{n}_\textrm{QD}
-
\lambda \, \hat{n}_c \hat{n}_\textrm{QD} .
\end{equation}
%
%
%
%%%%%%%%%%%%%%%%%%%%%%%%%%%%%%%%%%%%
%
%
Hereafter, we refer to the bonding state of the DQD as the effective ‘QD single-electron level’ throughout the paper.
Moreover, for definiteness, we choose for the rest of the discussion the bonding to antibonding energy transition in the DQD on the red side of the cavity mode, $\Omega<\omega_c$, so that we have $\lambda<0$.
As a consequence, we consider the initial level of the bonding state $\epsilon$ (with no photon in the cavity) to be below the chemical potential $(\epsilon<\mu)$ of the lead to which the DQD is laterally tunnel-coupled.
Alternatively, one can assume $\lambda>0$ and set $\epsilon>\mu$ for the initial state of the bonding state, see Fig.~\ref{fig:app_DQD_levels}.

\subsection{Tunneling dynamics with dispersive interaction}
\label{subsec:tunneling}

As mentioned above, the single QD level is tunnel-coupled to the lateral lead 
which is modeled as  a free Fermi gas. 
Then, starting  from the Bloch-Redfield-Wangness equation \cite{breuer2002theory} containing the tunneling between the lead
and the single QD level dispersively coupled to the detection mode, 
we trace out the degrees of freedom of the lead. 
We thus derive 
the following superoperator, valid in the single-photon regime 
%
%
%%%%%%%%%%%%%%%%%%%%%%%%%%%%%%%%%%%%
%
%   EQUATION N. 3
%
%
\begin{equation}
\begin{aligned}
&
\frac{\mathcal{L}_\textrm{QD} [\hat{\rho}]}{\Gamma_\textrm{QD}}  
= 
- f\left( \epsilon \right)
\left[ 
\left\{ \hat{\Pi}_0 \hat{d}\hat{d}^{\dagger} ; \hat{\rho} \right\}
-
2 \hat{\Pi}_0 \hat{d}^{\dagger}  \hat{\rho} \hat{d}\hat{\Pi}_0 
\right]
 \\
&
-  
f\left( \epsilon + \lvert\lambda\rvert \right) 
\left[ 
\left\{ \hat{\Pi}_1 \hat{d}\hat{d}^{\dagger} ; \hat{\rho} \right\}
-
2 \hat{\Pi}_1 \hat{d}^{\dagger} \hat{\rho} \hat{d} \hat{\Pi}_1
\right]
 \\
&
-  \left(1- f\left( \epsilon  \right) \right)
\left[ 
\left\{ \hat{\Pi}_0 \hat{d}^{\dagger} \hat{d}; \hat{\rho} \right\}
-
2 \hat{\Pi}_0 \hat{d} \hat{\rho} \hat{d}^{\dagger} \hat{\Pi}_0 
\right]
 \\
&
- \left(1- f\left( \epsilon + \lvert \lambda \rvert \right)  \right)
\left[ 
\left\{ \hat{\Pi}_1 \hat{d}^{\dagger} \hat{d}; \hat{\rho} \right\}
-
2 \hat{\Pi}_1 \hat{d} \hat{\rho} \hat{d}^{\dagger} \hat{\Pi}_1
\right]
 \\
&
+
\left[f\left( \epsilon  \right)
+
f\left( \epsilon + \lvert \lambda \rvert \right)
\right]
\left( 
\hat{\Pi}_0 \hat{d}^{\dagger} \hat{\rho} \hat{d} \hat{\Pi}_1 
+
\hat{\Pi}_1 \hat{d}^{\dagger} \hat{\rho} \hat{d} \hat{\Pi}_0
\right) 
 \\
&
+
\left[2-
f\left( \epsilon \right)
- 
f\left( \epsilon + \lvert \lambda \rvert \right)
\right]
\left( 
\hat{\Pi}_0 \hat{d}\hat{\rho} \hat{d}^{\dagger}  \hat{\Pi}_1 
+
\hat{\Pi}_1 \hat{d} \hat{\rho} \hat{d}^{\dagger} \hat{\Pi}_0
\right) 
\, . 
\label{eq:lead}
\end{aligned}
\end{equation}
%
%
%
%%%%%%%%%%%%%%%%%%%%%%%%%%%%%%%%%%%%
%
%  
%
%
Here, $\Gamma_\text{QD}$ is the tunneling rate of the QD describing the electron exchange  with the  lead and  the Fermi function $f( \varepsilon)=\left\{1 + \exp\left[ ( \varepsilon-\mu)/T\right]\right\}^{-1}$ with temperature $T$ and chemical potential $\mu$ of  the lead.
The two projectors 
$\hat{\Pi}_0 = 1 - \hat{n}_c $ 
and
$\hat{\Pi}_1 = \hat{n}_c$ 
define the projection onto the   Fock states of the detection mode with zero or one photon occupation. 
The first four terms of \cref{eq:lead} correspond to a Lindblad dissipator conditioned on photon occupation of the detection mode. The Fermi function appearing in the electron tunneling rate is adjusted accordingly,  depending on the photon occupation.
The remaining terms capture the coherences between the 0 and 1 photon subspace. 
In the limit $\lvert \lambda \rvert \rightarrow 0$, one recovers the Lindblad operator for a single QD level coupled to a Fermi reservoir. 
The full derivation of  \cref{eq:lead} 
is given in \cref{app:JumpOperators}.

In deriving Eq.~(\ref{eq:lead}), we assume that the QD electronic state is weakly tunnel-coupled to the lateral lead, so that the interaction can be treated perturbatively to first order (i.e., in the sequential tunneling regime).
Higher-order (cotunneling) terms, corresponding to quantum fluctuations of electron tunneling and leading to renormalization effects, are negligible. 
In our setup, when a photon enters the cavity, the QD is in a metastable state, assuming the single QD level  predominantly occupied. 
An electron can then tunnel out with the simple transition rate 
$\sim \Gamma_\text{QD}$ 
even in the low-temperature limit.
We argue that this perturbative Markovian approach where cotunneling and non-Markovian corrections are negligible 
is valid under the conditions: 
(i) 
The transparency of the barrier is low, i.e., the tunnel resistance is much larger than quantum resistance.
(ii) 
The electron conducting bandwidth of the contact is much larger than $\Gamma_\text{QD}$, 
(iii) The linewidth of the single electron level is smaller than the distance from the chemical potential, $\Gamma_{\textrm{QD}} \ll \left| \epsilon - \mu \right|$, which 
implies that statistical, fermionic correlations are irrelevant.

\subsection{Quantum cascade model}

In order to describe the incoming photon entering the detection cavity, we use the quantum cascade  approach  \cite{Carmichael1993,Gardiner1993}. 
It allows for a time-resolved description of a single-photon wavepacket impinging on the cavity 
and 
captures both the coherent evolution and the interaction-induced modifications of the quantum state during the scattering process.
This approach is based on a microscopic model involving an ancillary cavity mode 
(source mode) and a detection mode, both coupled to a chiral transmission line at zero temperature.

The source mode has frequency 
$\omega_\textrm{in}$ and bosonic operators, annihilation and creation, 
$
(\hat{a},
\hat{a}^{\dagger})
$, 
with occupation number
$\hat{n}_\textrm{in}  
=
\hat{a}^{\dagger}  \hat{a}^{\phantom{\dagger}}
$, 
and we set 
\begin{equation}
\hat{H}_{\textrm{in}} = 
\omega_\textrm{in} \hat{n}_\textrm{in} \, . 
\end{equation}
The decay of this source mode into the transmission line provides the incoming photon that hits the detector and sets its spectral linewidth.

After tracing out the transmission line, one obtains an effective 
dissipative, nonreciprocal interaction between the detection and source mode.
The corresponding   superoperator   involving the two modes' operators reads \cite{Carmichael1993,Gardiner1993}
%
%
%
%%%%%%%%%%%%%%%%%%%%%%%%%%%%%%%%%%%%
%
%   EQUATION N. 5
%
%
\begin{equation}
\begin{aligned}
\mathcal{L}_\textrm{cas} [\hat{\rho}] 
=& 
-\gamma_{\textrm{in}}
\left[ \left\{\hat{a}^{\dagger} \hat{a} ; \hat{\rho} \right\} - 2 \hat{a} \hat{\rho} \hat{a}^{\dagger} \right] 
 \\
&
-\gamma_{c}
\left[ \left\{ \hat{b}^{\dagger} \hat{b} ; \hat{\rho} \right\} - 2 \hat{b} \hat{\rho} \hat{b}^{\dagger} \right] 
\\
&
-2\sqrt{\gamma_{\textrm{in}} \gamma_c} 
\left[ 
\hat{b}^{\dagger} \hat{a} \hat{\rho} 
+ 
\hat{\rho} \hat{a}^{\dagger} \hat{b} 
- \hat{b} \hat{\rho} \hat{a}^{\dagger }
- \hat{a} \hat{\rho} \hat{b}^{\dagger }
\right] ,
\label{eq:cascade}
\end{aligned}
\end{equation}
%
%
%
%%%%%%%%%%%%%%%%%%%%%%%%%%%%%%%%%%%%
%
%   
%
%
%
where $\{\cdot;\cdot\}$ denotes the anticommutator.
The parameter $\gamma_{\textrm{in}}$ 
represents the bandwidth of the source whereas 
$\gamma_{c}$ corresponds the photon losses of the detection mode.
Although the derivation of Eq.~(\ref{eq:cascade}) was given in \cite{Carmichael1993} and in \cite{Gardiner1993}, 
for completeness, we provide another, independent derivation in 
\cref{app:cascade} starting from a Bloch-Redfield-Wangness equation \cite{breuer2002theory} 
containing the coupling between the two modes and the chiral line. 
Note that it is also possible to rewrite the dissipative interaction of
Eq.~\eqref{eq:cascade} in Lindblad form using jump operators given by a linear combination of the bosonic operators and introducing an effective coherent coupling between the 
source and the detection mode \cite{Carmichael1993}.

\subsection{Equation for the density matrix}

Combining the dissipators for electronic and photonic part,  the full time evolution of the composite system is governed by the master equation for the  density matrix
%
%
%
%%%%%%%%%%%%%%%%%%%%%%%%%%%%%%%%%%%%
%
%   EQUATION N. 6
%
%
\begin{align}
\frac{d \hat{\rho}}{dt} = & - i \left[H_S + \hat{H}_{\textrm{in}} ;\hat{\rho}\right]
%    +\mathcal{L}_\textrm{cas}\left[\hat{\rho} \right]
    +\mathcal{L}_\textrm{QD}\left[\hat{\rho} \right] +\mathcal{L}_\textrm{cas} [\hat{\rho}]  
    \, ,
    \label{eq:Lindblad}
\end{align}
%
%
%
%%%%%%%%%%%%%%%%%%%%%%%%%%%%%%%%%%%%
%
%   
%
%
where $\left[\cdot;\cdot\right]$ denotes the commutator. 
Then, \cref{eq:Lindblad} generates a set of coupled and closed linear equations for the matrix elements (see \cref{app:EoM}), which can be solved  exactly.

\section{Dispersive measurement and backaction} 
\label{sec:QND}
The goal is to detect a single photon by monitoring the alteration of the QD occupation 
in response to the presence of the photon in the cavity.
The dispersive nature of the coupling between the detection cavity mode and the QD ensures that the photon is not destroyed by the measurement process. 

Fixing the chemical potential of the lead $\mu=0$, 
we assume the system is initialized as follows.
The QD level is set to $\epsilon = -\lvert \lambda \rvert/2$ 
and the QD is in its stationary state with occupation probability $p^{(1)}(t = 0) = p^{(1)}_s$ and  
$p^{(0)}(t = 0) = p^{(0)}_s$. 
The subscript $s$ denotes the stationary state which is given by
$p^{(0)}_s  =   1-f\left( \epsilon \right)$
and
$p^{(1)}_s =  f\left(  \epsilon \right)$
for a given temperature $T$ of the lead.
 In the limit $T\to0$, the QD is fully occupied and $p^{(1)}_s=1$.  
 The detection mode with frequency $\omega_c$ is empty, $\langle \hat{n}_c(t=0)\rangle = 0$, whereas  the source mode is occupied, $\langle \hat{n}_\textrm{in}(t~=~0)\rangle~=~1$.
 The source mode 
 is on resonance with the cavity mode when its  frequency $\omega_\textrm{in}$
 is set to
$\omega_c + \lvert \lambda \rvert$, as the QD is initially occupied.
 If now the incoming photon occupies the detection mode, the QD level  shifts $\epsilon   \to \epsilon  + \lvert \lambda \rvert = \lvert \lambda \rvert/2$. 
 With an energy above the Fermi level of the lead, the electron on the dot can leave on a time scale set by $1/\Gamma_\textrm{QD}$.

\subsection{Time evolution}
\begin{figure}
    \centering
\includegraphics[width = \linewidth]{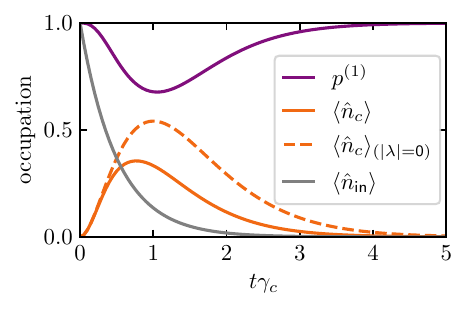}
    \caption{    
    Example of  time evolution of the QD occupation $p^{(1)}$ (purple), the source mode  $\langle\hat{n}_\textrm{in}\rangle$ (gray), and the detection mode $\langle\hat{n}_c\rangle$ (solid orange line) for $\omega_\textrm{in} = \omega_c +\lvert \lambda \rvert$, $\Gamma_\textrm{QD} = 4\gamma_c$, $\epsilon = -\lvert \lambda \rvert/2$, $\gamma_\textrm{in} = \gamma_c = 0.01 \omega_c$, 
    $\gamma_c  = 0.1 \, \lvert \lambda \rvert $, and $T = 0.01 \,  \lvert  \lambda \rvert$.
To show the effect of the  backaction due to the dispersive interaction, which always leads to 
a decrease of $\langle\hat{n}_c\rangle$, the detection mode occupation for $ \lambda  = 0$, $\langle\hat{n}_c\rangle_{( \lambda = 0)}$, is plotted for comparison (orange dashed line).
}
    \label{fig:evolution}
\end{figure}

The time evolution of the  system can be obtained  by solving \cref{eq:Lindblad}. 
The transient dynamics of the relevant observables, i.e., the QD probabilities $p^{(1)}(t)$ ($p^{(0)}(t)=1-p^{(1)}(t)$),  the occupation of the detection mode $\langle \hat{n}_c(t)\rangle$ and source mode $\langle\hat{n}_\textrm{in}(t)\rangle$ are shown in \cref{fig:evolution} 
for the initial conditions discussed above.

The average photon occupation of the source mode $\langle \hat{n}_\textrm{in}(t) \rangle$ decays exponentially on a time scale $\sim 1/ \gamma_\textrm{in}$. 
Due to the unidirectional interaction between 
source and detection mode, $\langle \hat{n}_\textrm{in}(t) \rangle$ is independent of all other system parameters 
\begin{align}
  \langle \hat{n}_\textrm{in}(t) \rangle = e^{-2\gamma_\textrm{in}t}.
  \label{eq:ana_nin}
\end{align}
The evolution of the average occupation of the detection mode, in the limit  $\lvert \lambda \rvert = 0$, when its evolution fully decouples from the QD, is given by
\begin{equation}
    \begin{aligned}
         {\langle \hat{n}_c(t)\rangle}_{( \lvert\lambda \rvert = 0)}  = 
         & \frac{8 \gamma_\textrm{in}\gamma_c} {\left(\gamma_\textrm{in} - \gamma_c\right)^2 + \Delta \omega^2} e^{-2\left(\gamma_\textrm{in}+\gamma_c\right)t}\\
     &
     \left\{\cosh\left[\left(\gamma_\textrm{in} - \gamma_c\right) t\right] - \cos\left(\Delta\omega t\right)\right\},
      \label{eq:ana_nc_lambda0}
    \end{aligned}
\end{equation}
with the detuning $\Delta \omega = \omega_\textrm{in} - \omega_c$. 
The initially empty cavity becomes populated with increasing time. 
The maximal occupation has an upper bound 
$n_\textrm{max}^{(0)} = 4/e^2$
for 
$\Delta\omega=0$ 
and  
$\gamma_\textrm{in}~=~\gamma_c$.
Then, for long times $t\gg 1/\gamma_c$, $\langle \hat{n}_c(t)\rangle$ decays exponentially.

In presence of the dispersive interaction with the QD
 $\lvert \lambda \rvert \neq0$, $\langle \hat{n}_c(t)\rangle$ 
 has a similar, non-monotonic behavior.  
The full evolution depends now on all system parameters. 
 Further analytical expressions are given 
in \cref{app:EoM}.
 The crucial general feature is that, overall, 
 the probability of the detection mode occupation is suppressed in presence of the dispersive 
 coupling with the QD.
We will analyze this measurement-induced backaction 
in the next section.

The evolution of the QD occupation is correlated with the one of the detection mode 
 with 
 $p^{(0)}(t) = 1 - p^{(1)}(t)$ 
 behaving  
 similarly to $\langle \hat{n}_c(t) \rangle$. 
 The electron in the QD level can tunnel out when the photon is in the cavity. 
 Once the photon left the cavity, the state of QD level relaxes to the initial state which, in the limit of low temperature $T\ll \lvert \lambda \rvert$, corresponds to the occupied state, as shown in Fig.~\ref{fig:evolution}.

%%%%%%%%%%%%%%%%%%%%%%%%%%%%%%%%%%%%%%%%%%%%%%%%%%%%%%%%%%%%%%
%
%
%   FIGURE N. 2
%
%

% 
%
%
%%%%%%%%%%%%%%%%%%%%%%%%%%%%%%%%%%%%%%%%%%%%%%%%%%%%%%%%%%%%%%

\subsection{Backaction of the dispersive measurement} 
\label{sec:backaction}
As shown in Fig.~\ref{fig:evolution}, we observe a backaction effect due to the dispersive coupling with the QD: The photon occupation of the detection mode
is significantly reduced compared to the uncoupled case.
This can be explained as a suppression of the coherence in the  transfer of the photon into the detection cavity.

To  analyze this result in detail, it is useful to analyze the behavior of the density matrix components  which we define as 
\begin{align}
    \rho^{c\tilde{c}}_{n_s n_d, \tilde{n}_s \tilde{n}_d} = \langle n_s,n_d,c\lvert\hat{\rho}\rvert 
    \tilde{n}_s, \tilde{n}_d, \tilde{c}\rangle.
\end{align}
Here, $c,\tilde{c} = 0,1$ denote the occupation of the QD, $n_s,\tilde{n}_s = 0,1$ is the occupation of the source mode and  $n_d,\tilde{n}_d = 0,1$ the number of photons in the detection mode. 
The equations of motion for each density matrix component are given in \cref{app:EoM}.

The initially occupied source mode $\rho_{10,10}^{c\tilde{c}}$ 
acts as source term in the equations of motion  for 
 $\rho^{c\tilde{c}}_{10,01}$ (off-diagonal in the photon subspace).
The latter is  itself  coupled to the occupation of the detection mode $\rho_{01,01}^{c\tilde{c}}$, such that we end up with sequential coupled equations for the components,
\begin{align}
    \rho_{10,10}^{c\tilde{c}} \quad \longrightarrow \quad \rho_{10,01}^{c\tilde{c}} \quad \longrightarrow \quad \rho_{01,01}^{c\tilde{c}}.
\end{align}
A reduction of $\rho^{c\tilde{c}}_{10,01}$ translates into a decrease of $\rho^{c\tilde{c}}_{01,01}$, therefore reducing the probability of absorption in the detection mode.
Thus, to explain the backaction, we focus hereafter on the behavior of the density matrix components $\rho^{00}_{10,01}$ and 
$\rho^{11}_{10,01}$ which are diagonal in charge, and off-diagonal in the photon subspace. 
They capture the coherence of the photonic part of the system playing a crucial role for the photon injection.

To qualitatively capture  the main features, we consider
 the limit of $T\to0$, where  
\begin{equation}
    \begin{aligned}
        \dot{\rho}_{10,01}^{00}  = & -2\sqrt{\gamma_\textrm{in}\gamma_c} \rho_{10,10}^{00} +\frac{\Gamma_\textrm{QD}}{2} \rho_{10,01}^{11} \\
        & -\left[{\frac{\Gamma_\textrm{QD}}{2} + \gamma_\textrm{in}+\gamma_c + i\Delta \omega}\right]\rho_{10,01}^{00}, \\
        \dot{\rho}_{10,01}^{11}  = & -2\sqrt{\gamma_\textrm{in}\gamma_c} \rho_{10,10}^{11} +\frac{\Gamma_\textrm{QD}}{2} \rho_{10,01}^{00} \\
        & -\left[\frac{\Gamma_\textrm{QD}}{2} + \gamma_\textrm{in}+\gamma_c + i
        \left( \Delta \omega  - \lvert \lambda \rvert \right)\right]\rho_{10,01}^{11}.
        \label{eq:eom_detectionmode_MT}
    \end{aligned}
\end{equation}
Setting $\lvert \lambda \rvert = 0$, the evolution of the QD and photon modes decouples. 
In this case \cref{eq:eom_detectionmode_MT} admit factorized solutions 
and one can consider 
the single equation of motion for the photonic off-diagonal component $\rho_{10,01} = \rho_{10,01}^{00} + \rho_{10,01}^{11}$ (fully independent of $\Gamma_\textrm{QD}$). 
Increasing the detuning $\Delta \omega$ leads to a decrease of $\rho_{10,01}$.

For $\lvert \lambda \rvert\neq 0$, in particular setting $\Delta \omega = \lvert \lambda \rvert$ in Eq.~(\ref{eq:eom_detectionmode_MT}),
we see that 
the evolution of $\rho_{10,01}^{11}$ is independent of $\lvert \lambda \rvert$, since 
the resonance condition between the detection mode and the incoming photon is fulfilled when the dot is occupied.  
In contrast, for the $\rho_{10,01}^{00}$ with $\lvert \lambda \rvert=\Delta\omega$, the dispersive coupling acts similarly to a detuning 
since
the  frequency of the detection mode is no longer on resonance with the incoming photon when the dot is empty.  
Consequently, increasing $\lvert \lambda \rvert$ has, qualitatively speaking, the same effect of reducing the cavity absorption probability as increasing the $\Delta \omega$ in the uncoupled case. 
This effect is significant when photon induced detuning 
is comparable to the linewidth  $\lvert \lambda \rvert \gtrsim \gamma_{c}$.

Another important observation is that the detection mode occupation does not vanish in the limit $\lvert \lambda \rvert \to \infty$, since there is always a finite probability that the electron does not tunnel to the lead for finite values of $\Gamma_\textrm{QD}$, (i.e., $\rho_{10,01}^{11}$ is independent of $\lvert \lambda \rvert$) and, in this case, the photon frequency remains unperturbed. 
In fact, when the tunneling process is slow (i.e., the tunneling rate is low $\Gamma_{\textrm{QD}}\ll\gamma_c$), the QD single-level does not have enough time to transfer its electron to the lead and therefore remains occupied even if the photon stays inside the cavity.
In this case, the photon cannot be detected.
Conversely, when tunneling is fast $\Gamma_{\textrm{QD}}\gg\gamma_c$, the electron quickly jumps to the lead, shifting the cavity mode out of resonance.
This possibility of rapid measurement significantly 
destroys coherence for 
$\lvert \lambda \rvert \gg \gamma_c$, 
lowering the probability that a photon will enter the cavity.

All the qualitative features discussed so far for  
$\langle \hat{n}_c(t)\rangle$ 
are consistent with the full solution which is stated in \cref{eq:app-nc_t}.

%%%%%%%%%%%%%%%%%%%%%%%%%%%%%%%%%%%%%%%%%%%%%%%%%%%%%%%%%%%%%%
%
%
%   FIGURE N.3
%
%
\begin{figure*}[t!]
    \centering
\includegraphics[width = 18cm]{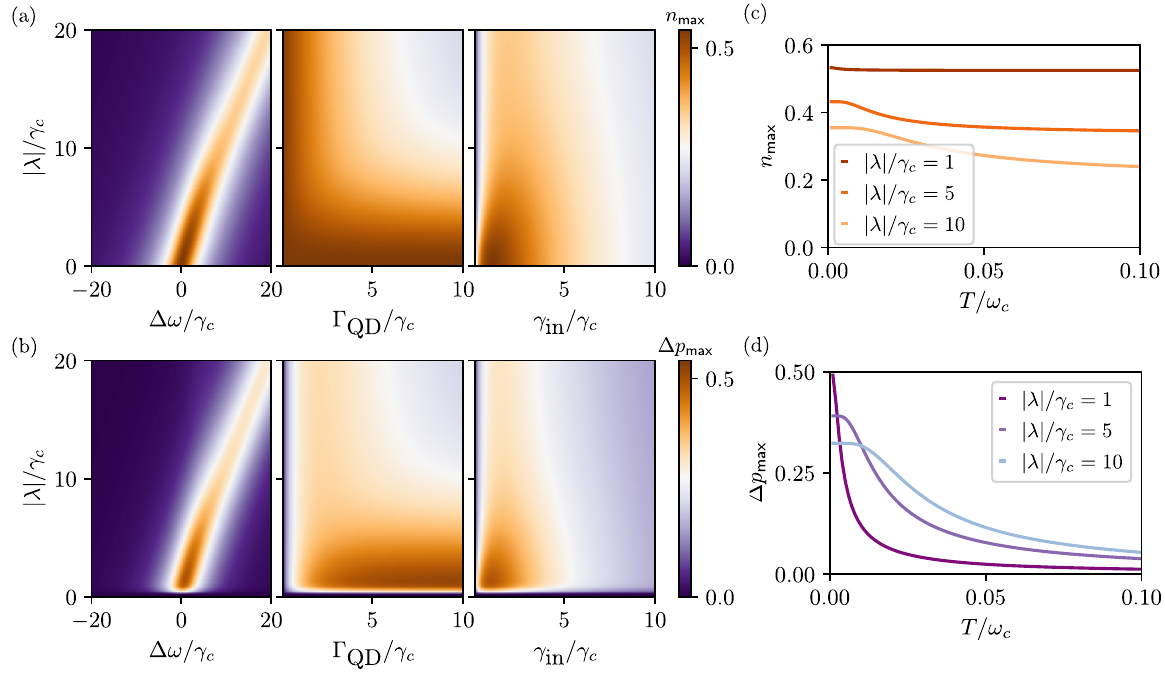}
    \caption{(a) $n_\textrm{max}$ 
    as function of $\lvert \lambda \rvert$ and system parameters 
    $\Delta \omega$ (left), $\Gamma_\textrm{QD}$ (middle), and $\gamma_\textrm{in}$ (right). 
    (b) $\Delta p_\textrm{max}$ as a function of the same  parameters as in (a). 
    (c) $T$ dependence of  
    $n_\textrm{max}$ and  (d) $\Delta p_\textrm{max}$ along cuts of constant $\lvert \lambda \rvert$.
    Constant parameters in all subfigures are set to 
    $T = 0.1 \gamma_c$, $\epsilon = -\lvert \lambda \rvert/2$ and  $\gamma_c = \gamma_\textrm{in} = 0.01 \omega_c$, $\Gamma_\textrm{QD} = 4\gamma_c$, $\Delta \omega = \lvert \lambda \rvert$. 
    }
    \label{fig:max_occupation}
\end{figure*}
%
%
%
%
%
%%%%%%%%%%%%%%%%%%%%%%%%%%%%%%%%%%%%%%%%%%%%%%%%%%%%%%%%%%%%%%

\subsection{Maximum photon occupation and QD depletion} 
\label{sec:max_occupation}

As the incoming photon only occupies the detection mode for a finite amount of time, the dispersive measurement relies on a transient effect. 
To summarize the dependence on various parameters, in the following 
we show:   
(i) 
The maximal detection mode occupation
$n_\textrm{max} = \textrm{max}_t  \langle \hat{n}_{c} (t)\rangle$, and (ii)
the maximal change in QD occupation over all times 
$t$, 
$
\Delta p_\textrm{max}~=~\textrm{max}_t
\left[
p^{(0)}(t) - p^{(0)}(t=0) 
\right]$, 
see \cref{fig:max_occupation}. 
In the following, we analyze for these two quantities the interplay between the different parameters, namely the detuning $\Delta \omega$, the QD rate $\Gamma_\textrm{QD}$, and the bandwidth of the incoming photon pulse $\gamma_\textrm{in}$, for given ranges of the dispersive coupling $\lvert \lambda \rvert$ and the temperature $T$.

\subsubsection{Dependence on the detuning $\Delta\omega$}

The left panel in Fig.~\ref{fig:max_occupation}(a) illustrates   
the matching condition needed to maximize the occupation of the detection photon mode for an initially occupied QD at $T\ll\lvert \lambda \rvert$. As expected, for a given $\lvert \lambda \rvert$ this resonance occurs at $\Delta\omega=\lvert \lambda \rvert$.

As a function of $\lvert \lambda \rvert$, the general trend is, as mentioned before, that the height $n_\textrm{max}$ of the resonance peak decreases for increasing $\lvert \lambda \rvert$.
  The  reduction predominately depends on how $\lvert \lambda \rvert$ compares to $\gamma_{c}$: for small enough $\lvert \lambda \rvert$ the shift of the cavity's resonance frequency between $\omega_c$ and $\omega_c + \lvert \lambda \rvert$ due to coupling to the QD is within the line width of detection mode.
 As $\lvert \lambda \rvert$ increases to $\lvert \lambda \rvert \gg \gamma_{c}$, the dispersive frequency shift  becomes larger  than the line width and  
the coherence between the photon modes is suppressed as explained in the previous section.

The maximal detection mode occupation is bounded from above by  $n_\textrm{max}^{(0)}$.
The lower bound is attained in the limit $\lvert \lambda \rvert \to \infty$, remaining strictly positive for finite value of the QD lifetime $1/\Gamma_\textrm{QD}$.
As explained before, for $\lvert \lambda \rvert \rightarrow \infty$, the photon occupation does not vanish   due to the finite probability that the QD  remains occupied  without inducing a frequency shift of the cavity mode. 

Concerning the QD, shown in the left panel of Fig.~\ref{fig:max_occupation}(b), $\Delta p_\textrm{max}$ 
 mirrors the behavior of $n_\textrm{max}$.
 The strong correspondence shows that by monitoring the QD depletion we have access to the occupation of the detection mode.
Notice that, for $\lvert \lambda \rvert \ll T$, thermal fluctuations of the QD smear out $\Delta p_{\max}$ (see the discussion in 
sec.~\ref{subsubsec:temperature}).

\subsubsection{Dependence on the tunneling rate $\Gamma_\textrm{QD}$}

The central panel 
of 
\cref{fig:max_occupation}(a) 
shows 
$n_\textrm{max}$  
for varying $\Gamma_\textrm{QD}$ and $\lvert \lambda \rvert$, 
assuming the resonant condition $\Delta \omega= \lvert \lambda \rvert$.

Here, 
we observe a first regime in which $n_{\max}$ remains almost unperturbed by the dispersive measurement, $n_{\max} \approx n_{\max}^{(0)}$.
This occurs for
$\lvert \lambda \rvert \ll \gamma_c$,
when $n_{\max}$ is almost unaffected by
$\Gamma_{\textrm{QD}}$.
In this case, the dispersive frequency shift of the detection mode induced by the QD charge depletion is smaller than the photon linewidth.
As discussed in the previous section, the cavity occupation is thus unaffected by the charge dynamics and does not depend on $\Gamma_{\textrm{QD}}$.

The second regime occurs for 
$\lvert \lambda \rvert \gg \gamma_\textrm{c}$ where 
one recovers the expected behavior discussed earlier:
For  $\Gamma_\textrm{QD} < \gamma_ c$ 
the tunneling process is slow, the QD's occupation does not have enough time to switch within the photon lifetime in the cavity,  resulting in an effective non-measurement.
Thus the photon occupation remains almost unaffected by the dispersive interaction.
As $\Gamma_\textrm{QD}$ increases, $\Gamma_{\textrm{QD}} \gg \gamma_\textrm{c}$, 
the photon occupation  diminishes due to the  QD measurement.

The central panel 
of 
\cref{fig:max_occupation}(b)
shows 
$\Delta p_{\max}$
as a function of  $\Gamma_\textrm{QD}$  and $\lvert \lambda \rvert$, 
assuming the resonant condition $\Delta \omega= \lvert \lambda \rvert$.
The effect of the finite temperature 
 is again visible at  small value of coupling $\lvert \lambda \rvert < T$, 
 where thermal fluctuations of the QD occupation 
 are comparable to the QD depletion caused by the photon cavity occupation 
 (see the discussion in 
Sec.~\ref{subsubsec:temperature}).
Excluding this region affected by temperature, the behavior of $\Delta p_{\max}$
is  correlated with that of $n_{\max}$ shown in the central panel of \cref{fig:max_occupation}(a)
and we summarize it in two distinct regimes.

In the first regime, when $\lvert \lambda \rvert > \gamma_c$, the dispersive interaction can affect the photon occupation.
However, at slow tunneling rates, $\Gamma_{\textrm{QD}} < \gamma_c$, the charge switching is slow, and the charge depletion $\Delta p_{\max}$ remains small: the QD occupation has a low probability of switching, i.e., 
$\Delta p_{\max}=0 $ for $\Gamma_{\textrm{QD}}\rightarrow 0$,  
and the photon entering the cavity is almost unperturbed. 
At fast tunneling rates, $\Gamma_{\textrm{QD}} > \gamma_c$, the charge can tunnel readily, allowing the measurement to occur. This induces backaction, which reduces the photon occupation and, in turn, decreases the probability that the electronic level is shifted  to allow the electron to tunnel.
The interplay of these two effects leads to a nonmonotonic behavior of $\Delta p_{\max}$ as a function of $\Gamma_{\textrm{QD}}$ when $\lvert \lambda \rvert > \gamma_c$.

In the second regime, when $\lvert \lambda \rvert < \gamma_c$, the dispersive interaction cannot affect the photon occupation, 
as visible in the central panel of \cref{fig:max_occupation}(a).
At the same time, at sufficiently fast tunneling rates, $\Gamma_{\textrm{QD}} >\gamma_c$, 
one may argue that the charge can tunnel readily, allowing the measurement to occur, as visible in the central panel of \cref{fig:max_occupation}(b).
In other words, in this region, we may expect to have both:
(i) 
The probability of the photon  entering the cavity 
remains unperturbed by the coupling to the QD 
(i.e., no backaction) 
(ii) High internal efficiency, i.e., 
the probability to detect the photon assuming that it is in the cavity is high. 
Note however that in this region, defined by $T<\lvert \lambda \rvert < \gamma_c < \Gamma_\textrm{QD} $, which a priori should be the optimal regime, our approach can unfortunately provide  only a qualitative picture, 
since $|\epsilon-\mu| = \lvert \lambda \rvert/2$ is comparable to or smaller than $\Gamma_{\textrm{QD}}$ and non-Markovian effects can become significant (see discussion in Sec.~\ref{subsec:tunneling}).

\subsubsection{Dependence on the bandwidth  $\gamma_\textrm{in}$}

The right panels of \cref{fig:max_occupation}(a) and (b) show the dependence of 
$n_\textrm{max}$ and 
$\Delta p_\textrm{max}$ on  $\gamma_\textrm{in}$. 

Regarding the cavity, $n_\textrm{max}$ reaches its maximum around $\gamma_\textrm{in} = \gamma_c$,
as discussed earlier.
Decreasing or increasing $\gamma_\textrm{in}$ leads to smaller values of $n_\textrm{max}$.
For larger values of the dispersive interaction $\lvert \lambda \rvert >0$,
the photon occupation as a function of $\gamma_\textrm{in}$ exhibits the same qualitative nonmonotonic behavior; however, the peak value of $n_\textrm{max}$ is suppressed as $\lvert \lambda \rvert$ increases, as discussed above.

Overall, the behavior of $\Delta p_{\max}$ shown in the right panel  of \cref{fig:max_occupation}(b) reflects that of $n_{\max}$ in most of its features. As a main exception, a marked difference is again visible in the narrow region $\lvert \lambda \rvert < T$ where, as mentioned above, the finite temperature washes out the effect of the photon.

\subsubsection{Dependence on the temperature  $T$}
\label{subsubsec:temperature}

The $T$ dependence of $n_\textrm{max}$ and $\Delta p_\textrm{max}$ are shown in \cref{fig:max_occupation}(c) and (d). 
First, we observe that, for $\lvert \lambda \rvert\leq \gamma_c$, effect of the QD on the photon is negligible and  $n_\textrm{max}$ results to be almost independent of $T$.
For larger values of $\lvert \lambda \rvert > \gamma_c$, instead,
the backaction of the QD becomes significant, reducing $n_{\max}$ at $T=0$.
As the temperature increases, for $T\gg \lvert \lambda \rvert$, the initial thermal equilibrium probabilities $p^{(0)}_s$ and $p^{(1)}_s$ saturate to $1/2$, since we have fixed $|\epsilon-\mu|=\lvert \lambda \rvert/2$.
On the one hand, such strong thermal fluctuations of the QD occupation lead to a random detuning of the cavity mode, resulting in a significant %pronounced backaction effect that 
reduction of the photon occupation, see \cref{fig:max_occupation}(c).
On the other hand, %in this regime the QD is available for detection only for the limited time when it is occupied, which leads to a decrease in the average QD depletion associated with the photon occupation, see \cref{fig:max_occupation}(d).
the thermal broadening of the lead is large enough that the level shift due to a photon in the cavity has no effect on the fluctuating tunneling dynamics, which suppresses $\Delta p_{\max}$, see \cref{fig:max_occupation}(d).

\section{Detection efficiency} 
\label{sec:efficiency}

Charge detection to measure the QD population is   routinely implemented in current experiments, for instance,   via monitoring the impedance of quantum %
nanoconductors capacitively coupled to the measured QD.
A discussion of this and other experimental issues is given in \cref{sec:discussions}.
Here we provide quantitative estimates for the detection efficiency of our proposed scheme by using a phenomenological model of a charge detector.

This model is based on an additional two-level system coupled only to the QD, which gets excited when the QD population changes from full to empty. 
This process can be included in the master equation (\ref{eq:Lindblad}) by adding the dissipator
%
%
%
%%%%%%%%%%%%%%%%%%%%%%%%%%%%%%%%%%%%%%
%
%
%       EQUATION N.
%
%
\begin{equation}
\begin{aligned}
    \frac{\mathcal{L}_\textrm{det}[\hat{\rho}]}{\gamma_\textrm{det}} & = -  \left[\left\{\hat{\Pi}_0^\textrm{QD}\hat{\sigma}^-\hat{\sigma}^+;\hat{\rho}\right\} - 2 \hat{\Pi}_0^\textrm{QD} \hat{\sigma}^+ \hat{\rho}\hat{\sigma}^-\hat{\Pi}_0^\textrm{QD}\right],
\end{aligned}
\end{equation}
%
%
%
%
%%%%%%%%%%%%%%%%%%%%%%%%%%%%%%%%%%
%
%
%
where $\hat{\sigma}^+,\hat{\sigma}^-$ are the raising/lowering operators for the detector two-level system and $\gamma_\textrm{det}$ denotes the charge detector rate. 
The operator $\hat{\Pi}_0^\textrm{QD}= 1 - \hat{n}_\textrm{QD}$ projects onto the unoccupied QD state, so to describe that the detector only clicks when the QD is empty. 
Adding this dissipator modeling the detector has no effect on the  dynamics of the system of the QD and of the cavity mode.
In order to estimate the detection efficiency, we consider that the QD is initially in the occupied state $p^{(1)}(t~=~0)~=~1$.  This can be ensured by monitoring the QD’s occupation, see \cref{sec:discussions}. At the same time, the detection two-level system is also initialized in its ground state.

At zero temperature $T=0$, the long-time limit of the detector's occupation 
$\langle \hat{n}_\textrm{det}\rangle$
coincides with the efficiency of detecting the photon, as
%
%
%
%%%%%%%%%%%%%%%%%%%%%%%%%%%%%%%%%%%%%%%%%
%
%
%   EQUATION N. 
%
%
\begin{align}
    \lim_{t\to\infty}\langle \hat{n}_\textrm{det}(t)\rangle < 1,
\end{align}
%
%
%
%
%%%%%%%%%%%%%%%%%%%%%%%%%%%%%%%%%%%%%%%%
%
%
%
where  $\hat{n}_\textrm{det} = \hat{\sigma}^+ \hat{\sigma}^-$.
However, at finite $T$ the QD occupation can change due to thermal fluctuations and consequently $ \langle \hat{n}_\textrm{det}\rangle\to 1$ for $t \to \infty$, independently of the occupation of the detection mode. 
Indeed, even for the QD decoupled from the detection cavity mode,  $ \lambda =0$, 
 $\langle \hat{n}^{(\lvert \lambda \rvert  = 0)}_\textrm{det}\rangle\to 1$ for $t \to \infty$.
Here, 
$\langle \hat{n}^{( \lvert \lambda \rvert = 0)}_\textrm{det}\rangle$ denotes the detector's occupation arising from the thermal background associated to the dark count rate, 
namely the  electron tunneling rate from the occupied QD into the lead at finite temperature
\begin{equation}
p_\textrm{dark} = \Gamma_\textrm{QD} \left[1-f\left(-\lvert \lambda \rvert/2\right) \right] \, . 
\end{equation}
For $T \ll\lvert \lambda \rvert$, $p_\textrm{dark} $ is exponentially small.

\begin{figure}
    \centering
    \includegraphics[width = 1.0\linewidth]{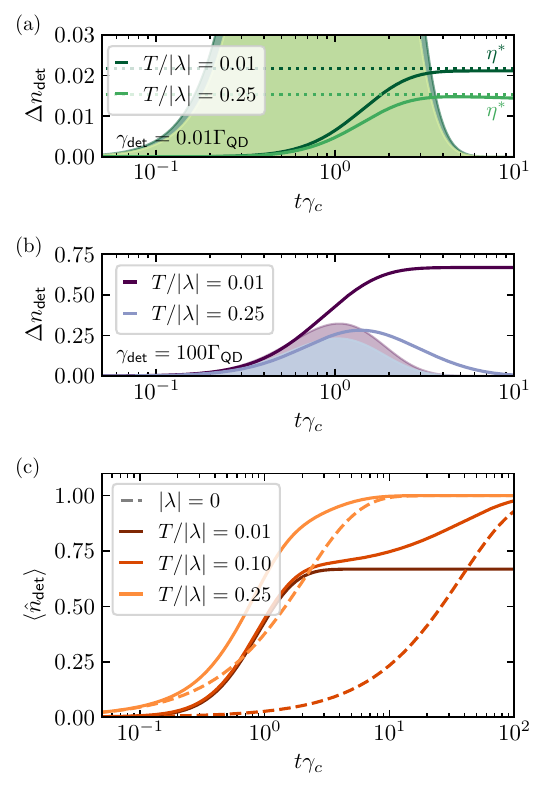}
    \caption{
    Full time evolution of $\Delta n_\textrm{det}(t)$ 
    for different temperatures, 
    for (a) $\gamma_\textrm{det} = 0.01 \Gamma_\textrm{QD}$ (solid lines)  
    and (b) $\gamma_\textrm{det} = 100 \Gamma_\textrm{QD}$ (solid lines).
    The evolution of $p^{(0)}(t)$ 
    is also shown for comparison (filled areas) for the corresponding choice of $T$. 
    In (a) the dotted lines correspond 
    to  $\eta^*$ (see text).
    (c)
    Full time evolution 
    $\langle \hat{n}_\textrm{det} \rangle$ (solid lines) 
    and 
     $\langle \hat{n}_\textrm{det}^{( \lvert \lambda \rvert =0)} \rangle$ 
     (dashed lines) 
    for different temperatures and for 
    $\gamma_{\textrm{det}}=100 \Gamma_\textrm{QD}$.
 Notice that the curve $\langle \hat{n}_\textrm{det}^{( \lvert \lambda \rvert  =0)} \rangle$ for $T = 0.01\,\lvert \lambda \rvert$ is so low that it is not visible.
    Parameters are set to 
    $\lvert \lambda \rvert=0.1 \, \omega_c$, 
    $\Gamma_\textrm{QD} = 4\gamma_c$,  
    $\gamma_c = \gamma_\textrm{in} = 0.01\omega_c$, $\epsilon=- \lvert \lambda \rvert/2$ 
    and  with initial QD state $p^{(1)}(t~=~0)~=~1$ 
    in all panels.
    }
\label{fig:efficiency}
\end{figure}

For a proper detector, one requires a very low value of the dark counts in the typical lifetime of the photon in the cavity and, at the same time, a fast tunneling rate to detect the photon  
\begin{equation}
p_\textrm{dark}
\ll
\gamma_{c} 
\ll 
\Gamma_\textrm{QD} \, .
\end{equation}
In this limit, taking into account  the thermal noise of the QD, we define
%
%
%
%%%%%%%%%%%%%%%%%%%%%%%%%%%%%%%%%%%%%%%%%%%%%%%%%
%
%
%   EQUATION N. 
%
%
\begin{align}
    \Delta n_\textrm{det}
    (t)= \langle \hat{n}_\textrm{det}(t)\rangle  - \langle \hat{n}^{( \lvert \lambda \rvert = 0)}_\textrm{det}(t)\rangle,
\end{align}
%
%
%
%%%%%%%%%%%%%%%%%%%%%%%%%%%%%%%%%%%%%%%%%
%
%
%
with  $\Delta n_\textrm{det} \to 0$ for $t \to \infty$.
Moreover, for sufficiently low dark count rate
(examples are shown in the dark curves in Fig.~\ref{fig:efficiency}(a-b)), a time-scale separation can occur: 
The effect of thermal dark counts remains small, and the behavior of $\Delta n_\textrm{det}$ is primarily determined by the behavior of the 
QD level occupation. 
In such a regime, $\Delta n_\textrm{det}(t)$  is (almost) constant for a time during which its value is directly related to the depletion of the QD level, and therefore correlated with  the photon occupation. 
In this regime, we define  
the detector efficiency as
%
%
%
%%%%%%%%%%%%%%%%%%%
%
%
%   EQUATION N.
%
%
\begin{align}
\eta = \max_{t} \Delta n_\textrm{det} (t).
\label{eqn:eta}
\end{align}

In ~\cref{fig:efficiency}(a) and (b), we show the time evolution of $\Delta n_\textrm{det}(t)$ for the limiting cases of a slow detector with $\gamma_\textrm{det} \ll \Gamma_\textrm{QD}$, and a fast detector with $\gamma_\textrm{det} \gg \Gamma_\textrm{QD}$, for two temperature choices: $T = 0.01\lvert \lambda \rvert$ and $T = 0.25\lvert \lambda \rvert$.

%%%%%%%%%%%%%%%%%%%%%%%%%%%%%%%%%%%%%%%%%%%%%%%%%%%%%%%%%%%%%%
%
%
%           FIGURE N.4
%
%
%
%
%
%
%
%%%%%%%%%%%%%%%%%%%%%%%%%%%%%%%%%%%%%%%%%%%%%%%%%%%%%%%%%%%%%%

In ~\cref{fig:efficiency}(a) we observe that 
 $\Delta  n_\textrm{det} (t)$ reaches a (quasi-) constant value for the selected temperature range. 
These values are reached almost immediately after the transient depletion of the QD level, and the probability $p^{(0)}(t)$ (plotted as filled areas in ~\cref{fig:efficiency}) returns to zero.
For low dark count and when $\gamma_\textrm{det}$ is the smallest rate, we obtain that the numerical value 
of $\eta$ is in a very good agreement with 
the linear response quantity 
$ \eta^* = \gamma_\textrm{det}\int_0^\infty \!\!\! dt \,  \Delta p^0(t)$ 
[dotted lines in \cref{fig:efficiency}(a)] .

In ~\cref{fig:efficiency}~(b), for $\gamma_\textrm{det} = 100 \Gamma_\textrm{QD}$, only 
in the low $T$ limit 
the time scales remain  well separated, the dark count rate  is low and 
the (quasi-) plateau is visible. 
The dark counts associated with the thermal 
noise becomes relevant already for $T=0.25\lvert \lambda\rvert $: 
$\Delta n_\textrm{det}$  does not show a plateau but simply a nonmonotonic behaviour.

In \cref{fig:efficiency}(c) we plot 
the full time evolution
$\langle \hat{n}_\textrm{det} \rangle$ (solid lines) 
compared to 
$\langle \hat{n}_\textrm{det}^{(\lvert  \lambda \rvert =0)} \rangle$ 
     (dashed lines) for $\gamma_{\textrm{det}} ~=~100 \Gamma_{\textrm{QD}}$ 
     for different temperatures. 
For high $T$, on the time scale $1/\gamma_c$,
the well-defined plateau that defines $\eta$
vanishes, as
$\langle \hat{n}_\textrm{det}(t) \rangle \to 1$ rapidly due to the increasing thermal noise.
For intermediate- and low-$T$, instead, the single photon yields a sharp increase up to the plateau and the dark counts are visible only at later times.

In summary, in the regime in which the dark count remains low,
(i.e., $\Delta n_\textrm{det}$  shows saturation  at finite time), \cref{eqn:eta} expresses  the efficiency for detecting a single photon.
Moreover, high detection efficiency can be reached in the limit of low temperature and for a sufficiently fast detector monitoring the state of  the QD coupled to the detection mode.

\section{Discussion and experimental challenges} 
\label{sec:discussions}

\subsection{Radio-frequency charge sensing}

To detect single-electron charges in nanostructures, charge sensors based on quantum point contacts (QPC) or QDs have long been established as powerful tools \cite{Field:1993,buks1998dephasing,Sprinzak:2002,Elzerman:2003}.
 The operation speed and sensitivity of such sensors can be greatly enhanced by embedding them in impedance-matching resonant circuits and monitoring the reflected radio-frequency (rf) signal, a technique known as rf reflectometry 
 \cite{schoelkopf1998radio,reilly2007fast,Barthel:2010}. 
 This approach is now routinely employed for fast, high-fidelity readout of charge and spin qubits in semiconductor QDs \cite{vigneau2023probing}.
For a broader tutorial on rf reflectometry principles, impedance matching, and backaction see Ref.~\cite{vigneau2023probing}.

An alternative QD charge detection scheme is gate-based reflectometry, where one of the confinement gates itself forms the capacitive element of the resonator and senses the QD’s quantum capacitance 
\cite{colless2013dispersive}; 
this eliminates the need for a separate nearby sensor. 
Both methods increase the charge detection bandwidth from the kHz range of dc conductance measurements up to the MHz--GHz range and enable single-shot charge or spin readout on microsecond 
timescales \cite{colless2013dispersive,blumoff2022fast,geng2025high,zheng2019rapid}.

Using an rf single-electron transistor (rf-SET) embedded in a donor quantum-dot circuit and coupled to a resonant tank circuit operating at approximately 225~MHz, Geng \textit{et al.}~\cite{geng2025high} demonstrated a characteristic rise--fall time constant of the charge-detector current of about 50~ns at base temperature. 
From signal-to-noise ratio (SNR) extrapolations, they reported an effective sensor integration time of $8 \pm 1.5$~ns (corresponding to $\mathrm{SNR}=2$) at 20~mK, establishing one of the fastest and most sensitive rf charge-detection schemes realized to date. 
More recently, Swift \textit{et al.}~\cite{swift2025superinductor} demonstrated a fully integrated, CMOS-compatible rf-SET matched via a TiN superinductor, achieving minimum integration times as short as $t_\mathrm{min} \approx 1 \pm 0.3~\mathrm{ps}$ with an estimated rise time of tens of nanoseconds---representing over two orders of magnitude improvement compared to previous rf-SET implementations.

These advances show that rf charge sensors readily achieve the bandwidth ($\gamma_\mathrm{det}\sim\mathrm{MHz\!-\!GHz}$) and SNR needed to resolve the transient QD depletion $p^{(0)}(t)$ expected from a single photon in our scheme.

\subsection{Electron temperature and its impact on detector performance}

We now briefly discuss the parameter regime experimentally accessible in hybrid circuit QED architectures with QDs, focusing on values relevant to our setup. 
For typical cavity frequencies of $\omega_{c}/(2\pi) \sim 10~\mathrm{GHz}$, an interdot tunneling rate of $t/(2\pi) \sim 10~\mathrm{GHz}$, and a bare charge--photon coupling of $g/(2\pi) \sim 1~\mathrm{GHz}$--- values that are routinely achieved in state-of-the-art DQD--cavity hybrid devices~\cite{scarlino2022situ}--- the resulting dispersive coupling is $\lvert \lambda \rvert/(2\pi) \sim 130~\mathrm{MHz}$ (see~\cref{app:DispersiveRegime}). 
High-quality-factor resonators with linewidths $\gamma_{c}/(2\pi)$ below $1~\mathrm{MHz}$ have been demonstrated~\cite{Harvey-Collard:2020, Holman:2021-npj, Holman:2020-APL}, placing the system well within the regime where 
$\gamma_{c} < \Gamma_{\mathrm{QD}} < \lvert \lambda \rvert$, 
with a typical QD tunneling rate with the lead $\Gamma_{\mathrm{QD}}$ in the tens of MHz.

As discussed previously, maintaining $\lvert \lambda \rvert/T\gg1$ is  essential to preserve a large readout contrast.
The electron temperature $T$ of the device is a central parameter governing the efficiency of our single-photon detector.   
The estimated value of $ \lvert \lambda \rvert/(2\pi)$ corresponds to a temperature of about $\sim 6$ mK.
For semiconductor QDs, achieving and maintaining $T \!\lesssim\!10~\mathrm{mK}$ already represents a demanding but realistic goal.

In semiconductor nanostructures, the refrigerator base temperature ($10$--$20~\mathrm{mK}$) does not directly translate into the electronic temperature $T$, which typically saturates at $50$--$120~\mathrm{mK}$ due to residual   heating and the vanishing electron–phonon coupling.
Substantial efforts are therefore required to reduce the electron bath temperature below this level.
For instance, immersion of the device in liquid Helium, 
combined with extensive low-pass filtering and
thermal anchoring of all electrical leads, 
has routinely given $T$ of a few mK
as inferred from Coulomb-blockade and QD
thermometry \cite{xia2000ultra,samkharadze2011integrated,iftikhar2016primary,nicoli2019quantum,jones2020progress,camenzind2022hole}. 
Pushing further, immersion of GaAs
two-dimensional electron gases in liquid 3He cooled by
nuclear adiabatic demagnetisation of a copper stage has
achieved sub-mK electron temperature  \cite{sarsby2020500,levitin2022cooling}.

In summary, the electron temperature is a technological bottleneck for the proposed detector.  
The steady progress towards  millikelvin and sub-millikelvin regimes demonstrates that the thermal conditions required for efficient photon-to-charge conversion are within reach of contemporary cryogenic technology, provided that the device is designed to minimize parasitic heat leaks and to exploit strong  coupling between the cavity mode and the DQD.

\section{Conclusions} 
\label{sec:conclusions}

To summarize, we have theoretically explored a circuit QED platform for non-absorptive detection of a single propagating microwave photon, based on the dispersive coupling between a resonant cavity mode and a DQD connected to a single  electronic lead. 
The development of a theory based on a quantum cascade approach has allowed a time-resolved description of the single-photon wavepacket incident on the cavity.

Specifically, we show how the variation of the electron level occupation is highly sensitive to the photon occupation in the regime $\gamma_{c} < \Gamma_{\textrm{QD}}$ 
and $T  < \lvert \lambda \rvert$. 
Making use of a phenomenological model of the DQD  coupled to a charge detector, we have assessed the detection efficiency of the proposed set-up.
Our results demonstrate that dispersive interactions on hybrid circuit QED platforms can generate measurable changes in the electron level occupation that might be suitable for charge-based readout in the microwave regime. 

At the same time, our analysis highlights the crucial role of temperature, which remains one of the main experimental challenges for the proposed scheme. 
Interestingly, the theoretical framework ---valid in the Markovian limit $\Gamma_{\mathrm{QD}} < \lvert \lambda \rvert$ --- naturally leads to a measurement-induced backaction effect:
our regime implies $\lvert \lambda \rvert > \gamma_{c}$, i.e., the dispersive coupling exceeds the cavity linewidth, thereby influencing the mechanism by which photons enter the cavity.

A promising direction for future work is to explore alternative detection schemes that offer partial immunity to thermal fluctuations. 
Examples include approaches based on interference effects in tunneling transport~\cite{Matern2023} or architectures relying on interdot tunneling events in buried QDs states, where the dots are well isolated from the leads by strong tunnel barriers \cite{Huang:2024-high}. 
Moreover, investigating the full counting statistics of the proposed detection protocol~\cite{Landi2024} represents an interesting and natural continuation of this work.

%%%%%%%%%%%%%%%%%%%%%%%%%%%%%%%%%%%%%%%%%%%%%%%%%%%%%%%%%%%%%%%%%%%%%%%
%               Acknowledgements
%%%%%%%%%%%%%%%%%%%%%%%%%%%%%%%%%%%%%%%%%%%%%%%%%%%%%%%%%%%%%%%%%%%%%%%
%

\begin{acknowledgments}
We  acknowledge continuous stimulating discussions with Christian Johansen, Anna Minguzzi, Maxime Richard, and Asian Selvakumaran. 
S.M., A.B., I.C. and G.R. acknowledge financial support 
from Provincia Autonoma di Trento (PAT); from the Q@TN Initiative;
from the National Quantum Science and Technology Institute
through 
the PNRR MUR project under Grant PE0000023-NQSTI, 
co-funded by the European Union -- NextGeneration EU.
P.S. acknowledges support from the Swiss State Secretariat for Education, Research and Innovation (SERI) under contract number MB22.00081~/~REF-1131-52105 and the support from the NCCR SPIN, a National Centre of Competence in Research, funded by the Swiss National Science Foundation (SNSF) with grant number 225153.
P.S. also acknowledges support from the SNSF through the grants Ref. No. 200021\_200418~/~1 and Ref. No. 206021\_205335~/~1.

\end{acknowledgments}

%%%%%%%%%%%%%%%%%%%%%%%%%%%%%%%%%%%%%%%%%%%%%%%%%%%%%%%%%%%%%%%%%%%%%%%
%                   APPENDICES
%%%%%%%%%%%%%%%%%%%%%%%%%%%%%%%%%%%%%%%%%%%%%%%%%%%%%%%%%%%%%%%%%%%%%%%
\appendix

%%%%%%%%%%%%%%%%%%%%%%%%%%%%%%%%%%%%%%%%%%%%%%%%%%%%%%
%%%%%%%%%%%%%%%%%%%%%%%%%%%%%%%%%%%%%%%%%%%%%%%%%%%%%%
%
%
%	A P P E N D I X	        A
%

\section{Derivation of the dispersive interaction between the DQD and the resonator}
\label{app:DispersiveRegime}

\subsection{The DQD Hamitonian}
We consider a double quantum dot (DQD) coupled to a single cavity mode via a dipole interaction decribed by
\begin{equation}
\hat{H}_\textrm{QED} = \hat{H}_\textrm{DQD} +   \omega_c \hat{b}^{\dagger} \hat{b} 
+   g\left( \hat{n}_1 - \hat{n}_2  \right) \left( \hat{b} + \hat{b}^{\dagger} \right)
\end{equation}
with 
\begin{equation}
\hat{H}_\textrm{DQD} = 
\bar{\varepsilon} \left( \hat{n}_1 + \hat{n}_2 \right)
+
\frac{\delta}{2} \left( \hat{n}_1 - \hat{n}_2 \right)
-
t_{12} \left( 
\hat{d}_1^{\dagger} \hat{d}_2^{\phantom{g}} 
+
\hat{d}_2^{\dagger} \hat{d}_1^{\phantom{g}}  \right). 
\end{equation}
Here, $\hat{n}_j = d^\dagger_jd_j$ are the occupations of the first ($j=1$) and second ($j=2$) dot, with the corresponding fermionic creation/annihilation operators $d^\dagger_j,d$. Additionally, $\bar{\varepsilon}$ is the average energy of the two dot levels, $\delta$ the DQD detuning and $t_{12}$ is the tunnel coupling between the dots. 
Introducing the  eigenstates of $\hat{H}_\textrm{DQD}$, namely the delocalized state of the electron on the two sites, we cast  $\hat{H}_\textrm{DQD}$ in terms of the fermionic 
operators for the ground and the excited state with one electron occupation 
\begin{equation}
\hat{H}_\textrm{DQD} = 
\bar{\varepsilon} \left( \hat{n}_e + \hat{n}_g \right)
+
\frac{ \Omega}{2}
\left(
\hat{d}_e^{\dagger} \hat{d}_e^{\phantom{g}}
-
\hat{d}_g^{\dagger} \hat{d}_g^{\phantom{g}}
\right)
\end{equation}
with the energy splitting $  \Omega= \sqrt{\delta^2 +  4 t_{12}^2}$ 
and the fermionic operators 
$\hat{d}_e~=~\cos\frac{\theta}{2} \hat{d}_1~-~\sin\frac{\theta}{2} \hat{d}_2$ 
and 
$\hat{d}_g~=~\sin\frac{\theta}{2} \hat{d}_1~+~\cos\frac{\theta}{2} \hat{d}_2$ 
where $\tan\theta= 2t_{12}/\delta$. 
We consider the case $\delta=0$ such that the system is described by a  Rabi-like model  
\begin{align}
\hat{H}_\textrm{Rabi} & =  \hat{H}_0 + \hat{H}_\textrm{int} \\
\hat{H}_0 &= 
\bar{\varepsilon} \left( \hat{n}_g + \hat{n}_e \right)
+
  \omega_c \hat{b}^{\dagger} \hat{b} 
+
\frac{ \Omega}{2}
\left(
\hat{d}_e^{\dagger} \hat{d}_e^{\phantom{g}}
-
\hat{d}_g^{\dagger} \hat{d}_g^{\phantom{g}}
\right)
\\
\hat{H}_\textrm{int}&=
g \left( \hat{b} + \hat{b}^{\dagger} \right)
\left( \hat{n}_1 - \hat{n}_2 \right) 
=
  g \left( \hat{b} + \hat{b}^{\dagger} \right)
\left( 
\hat{\sigma}_{+} 
+
\hat{\sigma}_{-} 
\right) \, .
\end{align}
with $\hat{\sigma}_{-} = \hat{d}^{\dagger}_g \hat{d}_e^{\phantom{g}}$.
and  $\hat{\sigma}_{+} = \hat{d}^{\dagger}_e \hat{d}_g^{\phantom{g}}$.
In the limit of weak interaction and far from the resonance 
\begin{equation}
g\ll \left( \Omega, \omega_c\right) \,, 
\qquad 
g\ll \left| \Omega- \omega_c\right| \,,
\end{equation}
we can apply a unitary Schrieffer-Wolff  transformation to determine the effective (low-energy) Hamiltonian by decoupling weakly interacting subspaces.

With the condition $\hat{H}_\textrm{int} = - \left[\hat{H}_0 ; \hat{S} \right]$ one obtains to second order 
\begin{equation}
\hat{H}'
= 
e^{-\hat{S}} \hat{H}_\textrm{Rabi} e^{\hat{S}}
\simeq 
\hat{H}_0 + 
\frac{1}{2}
\left[ \hat{H}_\textrm{int}; \hat{S} \right].
\end{equation}
Calculating the matrix elements between the unperturbed states of the Hamiltonian $\hat{H}_0$, we find the operator 
\begin{equation}
\hat{S}  = 
\frac{g}{\Omega-\omega_c} 
\left( \hat{\sigma}_- \hat{b}^{\dagger} - \hat{\sigma}_{+} \hat{b}^{\phantom{g}} \right)
+
\frac{g}{\Omega+\omega_c} 
\left( \hat{\sigma}_- \hat{b}^{\phantom{g}}  - \hat{\sigma}_{+} \hat{b}^{\dagger} \right) ,
\end{equation}
and the transformed  Hamiltonian corresponds to 
\begin{equation}
\label{eq:H_Prime}
\hat{H}'
=
\hat{H}_0
+ 
\hat{H}_\textrm{dis}
+
\lambda \left( \frac{\omega_c}{\Omega} \right)
\hat{\sigma}_x
+
\lambda   
\left( \hat{b}^{\phantom{g}} + \hat{b}^{\dagger}\right) \hat{\sigma}_z
\end{equation}
with $\hat{\sigma}_x = \hat{\sigma}_{+} + \hat{\sigma}_{-}$ and  $\sigma_z=\hat{n}_e-\hat{n}_g$ and the dispersive interaction
\begin{align}
\hat{H}_\textrm{dis}
=
 \lambda
\Big[ \hat{n}_c \hat{\sigma}_z + \hat{n}_e \Big],
\end{align}
where \cite{Zueco:2009,Zhu:2013,Clemens:2020} 
\begin{align}
\lambda = \frac{2g^2\Omega}{\Omega^2- \omega_c^2}
\,\, 
\left[
\approx \frac{g^2}{\Omega-\omega_c}
\,\,\,\,\, \mbox{for} \,\,\, \left|\Omega-\omega_c  \right| \ll \left( \Omega, \omega_c \right) 
\right] \, .
\label{eq:lambda_gen}
\end{align}
The $\hat \sigma_z$ and $\hat n$ factors in dispersive coupling $\hat{H}_\textrm{dis}$ provide a shift in opposite directions for the bonding and antibonding states of the DQD, of magnitude proportional to the photon number. 
Notice that we are considering the full Rabi model, so
Eq.~(\ref{eq:lambda_gen}) includes both the Jaynes–Cummings dispersive term as well as the Bloch–Siegert term \cite{Zueco:2009,Zhu:2013,Clemens:2020}.

The additional terms in \cref{eq:H_Prime} describe further processes which are however negligible in our case.
The third term   proportional to the operator $\hat{\sigma}_x$ 
produces a mixing between the excited and the ground state that we can neglect for sufficiently large energy splitting of the electronic levels, i.e.,   $\Omega \gg  \lvert \lambda \rvert \left(\frac{\omega_c}{\Omega}\right)$.
Moreover the fourth term in \cref{eq:H_Prime} proportional to $\hat{\sigma}_z$ 
leads to a renormalization of the cavity mode ground state which can be included 
by applying an unitary polaronic transformation 
$\hat{U}_p =\exp\left( \frac{2g^2\Omega}{\Omega^2-\omega_c^2}  \hat{\sigma}_z \hat{b}^{\dagger} + \mbox{h.c.} \right)$.
The validity of the Schrieffer-Wolff expansion holds on the condition 
\begin{equation}
    \frac{g^2}{\left|\Omega \pm \omega_c\right|} \ll \left(\Omega,\omega_c\right) \, .
\end{equation}

In the derivation presented here we have not assumed 
that the resonator frequency and the energy splitting of the DQD levels are sufficiently close to neglect the counter rotating term of the Rabi Hamiltonian \cite{Zueco:2009,Zhu:2013,Clemens:2020}.
This means that the effective dispersive Hamiltonian is also valid when, 
for instance, the detuning $\Delta=\Omega-\omega_c$ is comparable to frequencies of the system 
$\Delta \sim \left( \Omega ,\omega_c \right)$.

For example, when $\omega_c= \Omega/2$, we have $ \lambda= 4g^2/3\omega_c$ which corresponds to $ \lambda  \sim 130$ MHz for $g\sim 1$ GHz \cite{scarlino2022situ}, as stated in the main text.
\subsection{Reduction to a single electron level}
\label{app:subsec-singlelevel}

\begin{figure}[t!]
    \centering
    \includegraphics[width=0.9\linewidth]{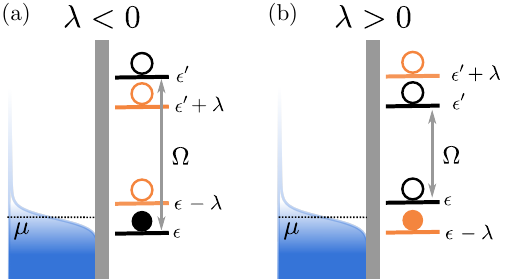}
    \caption{Energy levels of the DQD without photon (black) and in presence of a photon (orange). The splitting between the bonding state with energy $\epsilon$ and antibonding state with energy $\epsilon'$ is given by $\Omega$. In the presence of a photon $\Omega~\to~\Omega + 2\lambda$. 
    As explained in the text of Appendix \ref{app:subsec-singlelevel} the antibonding state always remains empty for our choice of parameters whereas the bonding state 
    is tuned such that (a) $ \epsilon < \mu$ if $\lambda <0$ 
    (i.e., the initial state is occupied) or 
     (b) 
     $ \epsilon > \mu$ if $\lambda >0$
     (i.e. the initial state is empty).}
    \label{fig:app_DQD_levels}
\end{figure}

The  splitting of the two electron levels  changes in presence of a photon in the cavity $\Omega \rightarrow \Omega +  2 \lambda $.
When coupled to a lead with chemical potential $\mu$, one can focus on a configuration for which the variation of the splitting is important only for the occupation of one electron level, see \cref{fig:app_DQD_levels} for a visualization of the energy levels used for photon detection.
We set $\epsilon=\bar{\varepsilon}-\Omega/2$
and 
$\epsilon'=\bar{\varepsilon}+\Omega/2$ as the bonding and the antibonding   levels.

One can impose, for example, that   
the antibonding state  of the DQD remains empty assuming   
$\epsilon' > \mu$ 
and 
$\epsilon'+ \lambda  > \mu$,
with 
$|\epsilon'  -\mu| \gg T$
and 
$|\epsilon' + \lambda-\mu| \gg T$.
This holds even for $|\lambda| \gg T$ since $\Omega \gg \lambda$, see \cref{fig:app_DQD_levels}. 

Simultaneously, 
we can have that the bonding state  occupation is strongly affected by the photon in the cavity.
For $\lambda>0$ [see \cref{fig:app_DQD_levels}(b)] we can assume
$\epsilon > \mu $ 
and 
$\epsilon  - \lambda <  \mu$.
For $\lambda<0$ [see of  \cref{fig:app_DQD_levels}(a)]  we can assume
$\epsilon < \mu $ 
and 
$\epsilon  - \lambda > \mu$.
In the latter case $(\lambda<0)$, the bonding state of the DQD is the single effective electronic level we consider in the main text.\\[1mm]

We summarize here the main points of the derivation presented in this Appendix: 
(i) 
we assume that the DQD detuning is zero $\delta=0$ 
so that the DQD's excitation is twice the interdot tunneling amplitude $t_{12}$; 
(ii) 
the coupling between the DQD and the cavity mode $g$ approaches the ultrastrong regime $g\approx 0.1 \,\omega_c$;
(iii) 
the frequency detuning between the cavity mode and the DQD $\left|\Omega-\omega_c \right|$ is much larger than $g$;
(iv) 
the chemical potential of the lead is set such that in the presence of a photon, the energy level of the bonding state is shifted above ($\lambda < 0)$ or below ($\lambda > 0)$ the chemical potential of the lead.

%%%%%%%%%%%%%%%%%%%

% Appendix   B

%%%%%%%%%%%%%%%%%%%
%

\section{Derivation of dissipative superoperator with the dispersive interaction} 
\label{app:JumpOperators}
We consider the  model Hamiltonian
\begin{equation}
\hat{H}_\textrm{tot}  = \hat{H}_s  + \hat{H}_\textrm{tun} + \hat{H}_\textrm{lead} .
\end{equation}
It is composed by the system of interest, a QD dispersively coupled to a resonator (equivalent to \cref{eq:H_s} in the main text)
\begin{equation}
\hat{H}_s  = \varepsilon_0 \hat{n}_d +   \omega_0 \hat{n}_c - \lambda \hat{n}_d \hat{n}_c, 
\end{equation}
where 
$\hat{n}_d = \hat{d}^{\dagger} \hat{d} $ is the occupation of the single-level QD and $\hat{d},\hat{d}^\dagger$ the fermionic annihilation and creation operator,
$\hat{n}_c = \hat{b}^{\dagger} \hat{b} $ is the occupation of the single mode resonator and $\hat{b},\hat{b}^\dagger$ the bosonic annihilation and creation operator, and
$\lambda$ the interaction strength of the dispersive coupling. We assume $\lambda < 0$ such that 
\begin{equation}
\hat{H}_s  = \varepsilon_0 \hat{n}_d +   \omega_0 \hat{n}_c + \lvert \lambda\rvert \hat{n}_d \hat{n}_c. 
\end{equation}

The QD is tunnel-coupled to a fermionic reservoir (the contact lead), which is described by 
\begin{align}
\hat{H}_\textrm{lead} &=  \sum_k \varepsilon_k \hat{c}^{\dagger}_k  \hat{c}^{\phantom{\dagger}}_k.
\end{align}
The lead is modeled as a free Fermi gas with the energy dispersion $\varepsilon_k$ and fermionic   annihilation and creation operators $\hat{c}_k,\hat{c}^\dagger_k$  at momentum $k$.
Its interaction with the QD is described by the tunneling Hamiltonian
\begin{equation}
\begin{aligned}
\hat{H}_\textrm{tun} &=  \left( \sum_k V_k \hat{c}_k \right)  \hat{d}^{\dagger} + \mbox{h.c}  \equiv \hat{F}\,\,  \hat{d}^{\dagger}  + \mbox{h.c},
\end{aligned}
\end{equation}
where $V_k$ is the tunneling amplitude. In the following we use the lead operator $F = \sum_k V_k\,c_k$.
\begin{widetext}

In the interaction picture a general operator is transformed as
$
\hat{O}_I(t) = e^{ i \hat{H}_0 t  } \hat{O} e^{- i \hat{H}_0 t}$, 
where $\hat{H}_0 = \hat{H_s} + \hat{H}_\textrm{lead}$.
For the lead (bath) and QD (system) operators we have
\begin{equation}
\begin{aligned}
\hat{F}_I(t) &=  \sum_k V_k  \hat{c}_k  e^{- i \varepsilon_k t} \, , \\
\hat{d}_I(t) &=  \hat{d} e^{- i \varepsilon_0 t  }\sum_{n=0}^{+\infty} \ket{n}\bra{n} e^{- i n \lvert \lambda \rvert t  } 
=  \hat{d} e^{- i \varepsilon_0 t  } \left[ \hat{\Pi}_0 + \hat{\Pi}_1 e^{- i  \lvert \lambda \rvert t  } +\dots \right]
\label{eq:dIt}
\end{aligned}
\end{equation}
where we have introduced the projector operator $\hat{\Pi}_n~=~\lvert n \rangle\langle n\rvert$  onto the Fock states of the resonator with $n$ photons. In the following we consider the the subspace with $n =0,1$. The relevant projectors for $n=0,1$ are then given by
\begin{equation}
\begin{aligned}
    \hat{\Pi}_0 & = 1 - \hat{n}_c, \\ 
        \hat{\Pi}_1 & = \hat{n}_c. 
\end{aligned}
\end{equation}

To derive a master equation  for the system's density matrix $\hat{\rho}_I = e^{ i \hat{H}_s t } \hat{\rho} e^{- i \hat{H}_s t }$ in the Markovian limit we start from the Bloch-Redfield-Wangness equation 
\begin{align}
\frac{d \hat{\rho}_I(t)  }{dt }
&=
- \int^{t}_{-\infty}\!\!\!\!\! dt'  \, \mbox{Tr}_L
\left[
\hat{H}_{\textrm{tun},I}(t) \hat{H}_{\textrm{tun},I}(t') \hat{\rho}_{L} \hat{\rho}_I(t)  
 - 
\hat{H}_{\textrm{tun},I}(t) \hat{\rho}_{L} \hat{\rho}_I(t)  \hat{H}_{\textrm{tun},I}(t')  
+ \mbox{H.c.} 
\right]
,
\label{app:eME-QD}
\end{align}
Where $\textrm{Tr}_L$ indicates the trace over the degrees of freedom of the lead. 
The Fermi reservoir is assumed at thermal equilibrium, such that its density matrix $\hat{\rho}_{L}~\propto~\exp\left( - \hat{H}_\textrm{lead} /  T\right) $.
Therefore the following correlation functions vanish
\begin{equation}
\langle  \hat{F}(t_1) \hat{F}(t_2) \rangle_L  = \langle \hat{F}^{\dagger}(t_1) \hat{F}^{\dagger}(t_2)\rangle_L  = 0,
\end{equation}
where we introduced 
${\langle \hat{F}(t) \hat{F}^{\dagger}(t') \rangle}_{L}~=~\mbox{Tr}_L\left[  \hat{F}(t) \hat{F}^{\dagger}(t')  \hat{\rho}_{L}  \right]$ for the correlation functions.
Then, \cref{app:eME-QD} simplifies to 
\begin{equation}
\frac{d \hat{\rho}_I(t)  }{dt }
=
I_1 + I_2 + I_3 + I_4 .
\label{eq:MEQD_terms}
\end{equation}
The individual terms are given by
\begin{align}
I_1 &= 
- \int^{t}_{-\infty}\!\!\!\!\! dt'  \,  R_h \left( t-t' \right)  \,\,\,  \hat{d}^{\dagger}_I(t) \hat{d}_I(t')  \,\, \hat{\rho}_I(t)+ \mbox{H.c.} 
 \, , \\
I_2 &= 
- \int^{t}_{-\infty}\!\!\!\!\! dt'  \, R^{*}_{e} \left( t-t' \right) \,\,\,  \hat{d}_I(t) \hat{d}^{\dagger}_I(t')  \,\, \hat{\rho}_I(t) + \mbox{H.c.}  \, , \\
I_3 &= 
\int^{t}_{-\infty}\!\!\!\!\! dt'  \, 
R_{e} \left( t-t' \right)  \,\,\,  \hat{d}^{\dagger}_I(t) \,\, \hat{\rho}_I(t) \,\, \hat{d}_I(t')  + \mbox{H.c.}  \, , \\
I_4 &= 
 \int^{t}_{-\infty}\!\!\!\!\! dt'  \, 
 R_h^* \left( t-t' \right)   \,\,\,  \hat{d}_I(t) \,\, \hat{\rho}_I(t)  \,\, \hat{d}^{\dagger}_I(t')  + \mbox{H.c.}  
\, ,
\end{align}
where we have introduced the correlators
\begin{align}
R_h \left( t-t' \right) = {\langle \hat{F}(t) \hat{F}^{\dagger}(t') \rangle}_{L}  &= \sum_k  
{\left| V_k \right|}^2 \left[ 1 - f\left(\varepsilon_k \right) \right]  e^{- i \varepsilon_k (t-t') } 
 \, , \\
R_e \left( t-t'\right) =
{\langle \hat{F}^{\dagger}(t') \hat{F}(t) \rangle}_{L}  &=  \sum_k {\left| V_k \right|}^2 f\left(\varepsilon_k \right)   e^{-  i \varepsilon_k (t-t')}
 \, , \\
 R_h^* \left( t-t' \right)
 =
{\langle \hat{F}(t') \hat{F}^{\dagger}(t) \rangle}_{L}   & =  \sum_k 
{\left| V_k \right|}^2\Big[ 1 - f\left(\varepsilon_k \right) \Big]    e^{  i \varepsilon_k (t-t') } 
 \, , \\
R_e^* \left( t-t' \right)
=
{\langle \hat{F}^{\dagger}(t) \hat{F}(t') \rangle}_{L}  
&= \sum_k  {\left| V_k \right|}^2
f\left(\varepsilon_k \right)   e^{ i \varepsilon_k (t-t') } 
\, ,
\end{align}
with the Fermi function $f(\varepsilon)$.  
Defining
\begin{equation}
\begin{aligned}
R_h\left( E \right) &=
 \int^{t}_{-\infty}\!\!\!\!\! dt'  
R_h \left( t-t' \right) \,\, e^{ i E \left( t-t' \right) }  = 
 \int^{t}_{-\infty}\!\!\!\!\! dt'  
\sum_k  {\left| V_k \right|}^2
\Big[ 1 - f\left(\varepsilon_k \right) \Big]  e^{ i \left( E- \varepsilon_k  \right)  \left( t-t' \right) } 
 \, , \\
R_e \left( E \right) &=
 \int^{t}_{-\infty}\!\!\!\!\! dt'  
R_e \left( t-t' \right) \,\, e^{ i E \left( t-t' \right) } 
 \int^{t}_{-\infty}\!\!\!\!\! dt'  
\sum_k 
{\left| V_k \right|}^2  
f\left(\varepsilon_k \right)   e^{ i \left( E - \varepsilon_k  \right)  \left( t-t' \right) } 
\end{aligned}
\end{equation}
and using the functional decomposition 
\begin{equation}
\begin{aligned}
 \int^{t}_{-\infty}\!\!\!\!\! dt'   e^{i \Omega \left( t-t' \right) }
=
\lim_{\eta=0} \int^{\infty}_0 \!\!\!\! d\tau  \,\,\, e^{i \Omega \tau } e^{-\eta \tau } 
=
\pi\delta\left( \Omega \right) +i \mbox{P}\left( \frac{1}{\Omega}\right) 
\, ,
\end{aligned}
\end{equation}
one obtains
\begin{align}
R_h\left( E \right) &=
 \Big[ 1 - f\left(E \right) \Big] \Gamma\left( E \right)  + i \,\, \mbox{Im} \left[  R_h\left( E \right) \right]
 \, , \\
R_e \left( E \right) &=
 f\left(E \right)  \Gamma\left( E \right)  + i \,\,\mbox{Im} \left[  R_e\left( E \right) \right]
 \, ,
 \end{align}
with
\begin{equation}
  \Gamma\left( E \right)  = \pi \sum_k {\left| V_k \right|}^2   \delta\left( \varepsilon_k - E  \right)
\, . 
\end{equation}
Hereafter we only consider the real part of the function $R_h\left( E \right)$ and $R_e \left( E \right)$ and neglect the imaginary part corresponding to a Lamb shift correction. 
Moreover, we assume that the tunneling rate is independent of the energy and we set $  \Gamma\left( E \right)   \equiv \Gamma_\textrm{QD}$. Then, the correlation functions are given by
\begin{align}
R_h\left( E \right) & 
=  \left[ 1 - f\left(E \right) \right]\Gamma_\textrm{QD}   , \\
R_e \left( E \right) &
 = f\left(E \right)  \Gamma_\textrm{QD}  
 \, .
 \end{align}
Using the time evolution of the QD operators in the single photon subspace defined in  \cref{eq:dIt} we obtain
\begin{equation}
\begin{aligned}
I_1 
& = 
- \!\! \int^{t}_{-\infty}\!\!\!\!\! dt'  \,  R_h \left( t-t' \right)   \hat{d}^{\dagger} \hat{d} e^{ i \varepsilon_0 (t-t')} \left[ \hat{\Pi}_0 +  \hat{\Pi}_1 e^{ i \lvert \lambda \rvert (t-t')}  \right]  \hat{\rho}_I(t) + \mbox{H.c.}  \\
 & = 
-  \left[ R_h \left( \varepsilon_0  \right) \hat{\Pi}_0  + R_h \left( \varepsilon_0 +\lvert \lambda \rvert  \right)   \hat{\Pi}_1   \right] \, \hat{d}^{\dagger} \hat{d}   \,\, \hat{\rho}_I(t)+ \mbox{H.c.} 
\nonumber \\
 & 
\simeq 
- \Gamma_\textrm{QD} 
\left\{ 
\left[ 
 \Big( 1 - f(\varepsilon_0) \Big)
\hat{\Pi}_0 
+
 \Big( 1 - f(\varepsilon_0+\lvert \lambda \rvert) \Big)   \hat{\Pi}_1   
 \right] \, \hat{d}^{\dagger} \hat{d}   ; \hat{\rho}_I(t) 
 \right\} \, .
\end{aligned}
\end{equation}
Similarly, we find for the remaining terms of \cref{eq:MEQD_terms}
\begin{equation}
\begin{aligned}
I_2
 & = \simeq 
- \Gamma_\textrm{QD} \left\{ \left[ 
 f(\epsilon)
\hat{\Pi}_0 
+
 f(\epsilon+\lvert \lambda \rvert)   \hat{\Pi}_1   \right] \, \hat{d}  \hat{d}^{\dagger}   ; \hat{\rho}_I(t) \right\},\\
I_3 & \simeq
\Gamma_\textrm{QD} 
\Big[ 
2  f(\epsilon) \hat{\Pi}_0  \hat{d}^{\dagger} \hat{\rho}_I(t) \hat{d} \, \hat{\Pi}_0 
+ 2  f(\epsilon+\lvert \lambda \rvert) \hat{\Pi}_1  \hat{d}^{\dagger} \hat{\rho}_I(t) \hat{d} \, \hat{\Pi}_1 \\
& +
\left(    f(\epsilon) +  f(\epsilon+\lvert \lambda \rvert) \right)
\left( 
 \hat{\Pi}_0  \hat{d}^{\dagger} \hat{\rho}_I(t) \hat{d} \, \hat{\Pi}_1 e^{- i \lvert \lambda \rvert t}
 +
 \hat{\Pi}_1  \hat{d}^{\dagger} \hat{\rho}_I(t) \hat{d} \, \hat{\Pi}_0 e^{ i \lvert \lambda \rvert t}
\right)
\Big] \\
I_4 
&
\simeq
\Gamma_\textrm{QD} 
\Big[ 
2 \left[1 - f(\epsilon) \right]  \hat{\Pi}_0  \hat{d} \hat{\rho}_I(t) \hat{d}^{\dagger} \, \hat{\Pi}_0  
+ 2  \left[1 -  f(\epsilon+\lvert \lambda \rvert) \right] \hat{\Pi}_1  \hat{d} \hat{\rho}_I(t) \hat{d}^{\dagger} \, \hat{\Pi}_1 \\
& +  
\left( 2-   f(\epsilon) -  f(\epsilon+\lvert \lambda \rvert)   \right) 
 \hat{\Pi}_0  \hat{d} \hat{\rho}_I(t) \hat{d}^{\dagger}  \, \hat{\Pi}_1 e^{ i \lvert \lambda \rvert t} 
\left( 2-   f(\epsilon) -  f(\epsilon+\lvert \lambda \rvert)   \right)
 \hat{\Pi}_1  \hat{d} \hat{\rho}_I(t) \hat{d}^{\dagger}  \, \hat{\Pi}_0 e^{- i \lvert \lambda \rvert t}
\Big] 
\label{eq:I2I3I4}
\end{aligned}
\end{equation}
The terms $I_3$ and $I_4$ contain an explicit time dependence. Such oscillating factors are often neglected using 
the secular approximation that holds if the oscillations are fast, namely $\lvert \lambda \rvert$ is much larger than the typical time scale evolution of the matrix elements of the density matrix, which here corresponds to $\Gamma_\textrm{QD}$. 
Here we do not apply the secular approximation as we consider the limit $\lvert \lambda \rvert\rightarrow 0$.

These oscillating factors  in $I_3$ and $I_4$ of Eq.~(\ref{eq:I2I3I4}) disappear when transforming the density matrix in \cref{eq:MEQD_terms} into the Schr\"odinger picture,
$
    \hat{\rho} \left(  t \right) =
e^{- i \hat{H}_s t } \hat{\rho}_I\left(  t \right)     e^{ i \hat{H}_s t }$. 
This leads to the master equation
\begin{align}
    \frac{d \hat{\rho} \left(  t \right) }{dt}
=
- i\left[ \hat{H}_S ;  \hat{\rho}(t)  \right]
+
\mathcal{L}_\textrm{QD}\left[\hat{\rho}\right],
\end{align}
where the Liouville superoperator $\mathcal{L}_\textrm{QD}$ is given by \cref{eq:lead} in the main text.

%%%%%%%%%%%%%%%%%%%%%%%%%%%%%%%%%%%%%%%%%%%%%%%%%%%%%%
%%%%%%%%%%%%%%%%%%%%%%%%%%%%%%%%%%%%%%%%%%%%%%%%%%%%%%
%
%
%	A P P E N D I X	        C
%
%
\section{Derivation of the cascade model}
\label{app:cascade}
The quantum cascade model was first proposed in Refs.\,\cite{Carmichael1993,Gardiner1993}.
Here we give an alternative derivation.
We first recover the equations of motion equivalent to
Ref.\,\cite{Gardiner1993} (interaction picture) and   
Ref.\,\cite{Carmichael1993} (Heisenberg picture) and then we derive the dissipative superoperator starting from the Bloch-Redfield-Wangness equation.

We consider the following model in which two localized modes are coupled to 
a chiral transmission line in which the photon propagates only in one direction, with momentum $k>0$,
\begin{equation}
\hat{H}_\textrm{cas} = 
\hat{H}_\textrm{cas}^{(0)} 
+
\hat{H}_\textrm{cas}^{(I)} 
\, , 
\end{equation}
where
\begin{align}
\hat{H}_\textrm{cas}^{(0)} 
&= 
 \omega_a \hat{a}^{\dagger} \hat{a}^{\phantom{g}}
+
 \omega_b \hat{b}^{\dagger} \hat{b}^{\phantom{g}}
+
\int^{\infty}_0\!\!\!\!\!\! d\omega \,  \omega \, \hat{a}_{\omega}^{\dagger}
\hat{a}_{\omega}^{\phantom{g}}
\, ,
\\
\hat{H}_\textrm{cas}^{(I)} 
&= 
g_a\left( \hat{a}^{\phantom{g}} + \hat{a}^{\dagger} \right) \hat{\phi}(x_a)
+
g_b\left( \hat{b}^{\phantom{g}} + \hat{b}^{\dagger} \right) \hat{\phi}(x_b)
\, . 
\end{align}
The field operator takes the general form
\begin{equation}
\hat{\phi}(x)
=
\int^{\infty}_0\!\!\!\! d\omega \, \phi\left(\omega \right)
\left[ \hat{a}_{\omega} e^{i\frac{\omega}{v} x } + \mbox{h.c.} \right]
\, ,
\end{equation}
with   $v$ the velocity of the light in the one dimensional line.
Using the rotating wave approximation, we write 
\begin{equation}
\hat{H}_\textrm{cas}^{(I)} 
= \int^{\infty}_0 \!\! d\omega \, \phi\left(\omega \right)
\left[  
g_a e^{-i\frac{\omega}{v} x_a}  \hat{a} 
+
g_b e^{-i\frac{\omega}{v} x_b}  \hat{b}  
\right]
\hat{a}_{\omega}^{\dagger}
+
 \mbox{H.c.}
\, .
\end{equation}

\subsection{Equations of motion}

By solving the equation of motion for the operator $\hat{a}_{\omega}$ in the Heisenberg picture, one finds 
\begin{align}
 \hat{a}_{\omega}(t)
= 
\hat{a}_{\omega}(t_0) e^{- i \omega (t-t_0)}
- i
\int^t_{t_0}
 dt' 
\phi(\omega) e^{- i \omega (t-t')}
\left[  
g_a e^{-i\frac{\omega}{v} x_a}  \hat{a}(t')
+
g_b e^{-i\frac{\omega}{v} x_b}  \hat{b}(t')
\right].
\end{align}
Then for the operator of the bosonic mode $\hat{a}$ the equation of motion reads   
\begin{align}
\frac{d \hat{a}(t)}{dt} =
- i\omega_a \hat{a}(t) 
-
g_a^2 \int^{t}_{t_0}\!\!\!\! dt'  K(t-t') \hat{a}(t')
- 
g_a g_b \int^{t}_{t_0} dt' K(t+\tau-t')\, \hat{b}(t')
-
 i g_a 
\int^{\infty}_0 d\omega 
\phi(\omega)\, e^{-i\omega(t-t_0) + i \frac{\omega}{v} x_a}
\hat{a}_{\omega} , 
\label{eq:a_t}
\end{align}
where we set $\tau = (x_b-x_a)/v > 0$ and the response function 
\begin{equation}
K(t-t') = 
\int^{+\infty}_0\!\!\!\!\!\! d\omega 
{\left| \phi(\omega) \right|}^2 \, e^{-i\omega(t-t')} \,.
\label{eq:K_t}
\end{equation}
For the function $K(t-t') $ we can disregard the imaginary part that lead to a weak renormalization of the eigenstates and of the spectrum. 
We also assume that $K(t-t') $ satisfies causality, i.e., 
$K=0$ for $t'>t$.
In the Markovian limit, the function $K(t-t') $ can be treated as 
a $\delta$-function provided that its characteristic time scale is faster than the typical time scale in the evolution of the operators $\hat{a}$ and $\hat{b}$. 
In this limit, we write 
\begin{equation}
K(t-t') \approx K_0\, \delta(t-t') \label{eq:K0delta},
\end{equation}
This implies that the third term appearing in \cref{eq:a_t}
vanishes since $\delta(t+\tau-t')$ the time $t+\tau$ is out of the integration range and one obtains 
\begin{equation}
\frac{d \hat{a}(t)}{dt} 
=
- i\omega_a \hat{a}(t) 
- \frac{\gamma_a}{2} \hat{a}(t) 
- \sqrt{\gamma_a} \hat{a}_{in}(t),
\label{eq:da/dt}
\end{equation}
where the damping coefficients (photon losses) are defined as
\begin{equation}
\gamma_a = \frac{1}{2} g_a^2 K_0 \, , \qquad
\gamma_b = \frac{1}{2}  g_b^2 K_0 \, ,  \label{eq:gagb}
\end{equation}
and the noise 
\begin{equation}
\sqrt{\gamma_a} \, \hat{a}_{in}(t)
=
-
 i g_a 
\int^{+\infty}_0\!\!\!\!\!\! d\omega 
\phi(\omega) e^{-i\omega(t-t_0) + i \omega x_a / v}
\hat{a}_{\omega},
\end{equation}
such that $[\hat{a}_{in}^{\phantom{g}}(t);\hat{a}_{in}^{\dagger}(t')] = \delta(t-t')$. 
We remark that we use the Ito convention $\int^{t}_{t_0}\!dt' \delta(t-t')= 1/2$.
In a similar way, one can write the equation of motion for the bosonic operator $\hat{b}$  which reads 
\begin{align}
\frac{d \hat{b}(t)}{dt} 
&
=
- i\omega_b \hat{b}(t) 
-
g_b^2 \int^{t}_{t_0}\!\!\!\! dt'  K(t-t') \hat{b}(t')
- 
g_a g_b \int^{t}_{t_0}\!\!\!\! dt' K(t-\tau-t') \hat{a}(t')
-
 i g_b
\int^{+\infty}_0\!\!\!\!\!\! d\omega 
\phi(\omega) e^{-i\omega(t-t_0) + i \omega x_b / v}
\hat{a}_{\omega}  \, .
\label{eq:b_t}
\end{align}
Notice that, in the third term of \cref{eq:b_t}, the time $\tau$ appears with the minus sign and therefore, in the Markovian limit, we obtain
\begin{equation}
\frac{d \hat{b}(t)}{dt} 
=
- i\omega_b \hat{b}(t) 
- \frac{\gamma_b}{2} \hat{b}(t) 
- \sqrt{\gamma_b} \hat{a}_{in}(t)
- \sqrt{\gamma_a\gamma_b} \hat{a}(t-\tau).
\label{eq:db/dt}
\end{equation}
Comparing \eqref{eq:da/dt} and \eqref{eq:db/dt}, it is evident that the resonant mode $\hat{b}$ is affected by the field of the resonant mode $\hat{a}$ but not viceversa.

\subsection{Dissipative operator in the master equation}

We switch to the interaction picture and we write 
\begin{equation}
\hat{H}_\textrm{cas}^{(I)} (t) = \int^{\infty}_0\!\!\!\!\! d\omega 
\left[
\hat{A}_{\omega}^{\dagger}(t) \hat{C}_{\omega}^{\phantom{g}}(t)
+ \mbox{H.c.} \right] \, , 
\end{equation}
with 
\begin{equation}
\hat{A}_{\omega}(t) 
=
\phi(\omega) e^{-i\omega t} \hat{a}_{\omega}  \, , 
\end{equation}
and
\begin{equation}
\hat{C}_{\omega}(t) =
g_a e^{-i\frac{\omega}{v} x_a}  e^{-i\omega_a t } \hat{a}
+
g_b e^{-i\frac{\omega}{v} x_b}  e^{-i\omega_b t } \hat{b}  \, . 
\end{equation}
Using the Born-Markov approximation, considering the coupling with the line as the interaction (perturbation), one has the equation 
for the density matrix of the cascade system $\hat{\rho}^{(c)} $ 
formed by the resonator $a$ and $b$, 
in the interaction picture
\begin{equation}
\hat{\rho}_I^{(c)}(t) = e^{ i \hat{H}^{(0)}_\textrm{cas} t } \hat{\rho}^{(c)} (t) e^{- i \hat{H}^{(0)}_\textrm{cas} t } \, .
\label{eq:rho_I}
\end{equation}
We start from the Bloch-Redfield-Wangness equation
\begin{align}
\frac{d} {dt }\hat{\rho}_I^{(c)}(t)  
=
-\int^{t}_{-\infty}\!\!\!\!\! dt'  \, \mbox{Tr}_\textrm{line}
\left[
\hat{H}_\textrm{cas}^{(I)}(t) \hat{H}_\textrm{cas}^{(I)}(t') \hat{\rho}_{\ell} \hat{\rho}_I^{(c)}(t)   - 
\hat{H}_\textrm{cas}^{(I)}(t) \hat{\rho}_{\ell} \hat{\rho}_I^{(c)}(t)  \hat{H}_\textrm{cas}^{(I)}(t')  
+ \mbox{H.c.} 
\right].
\label{eq:BRW-1}
\end{align}
The chiral photonic 1D line  is assumed to be at thermal equilibrium  $\hat{\rho}_{\ell}^{(th)}$. In particular we consider the 
zero temperature limit $T=0$ for the line, namely
$\left< \hat{a}_{\omega}^{\dagger} \hat{a}_{\omega}^{\phantom{g}} \right>~=~0$ whereas 
$\left< \hat{a}_{\omega}^{\phantom{g}} \hat{a}_{\omega' }^{\dagger}  \right>~=~\delta(\omega-\omega')$.
Then, for the non-squeezed vacuum, the following correlation function vanish 
\begin{equation}
\begin{aligned}
\mbox{Tr}_\textrm{line}\left[  \hat{A}_{\omega}(t_1) \hat{A}_{\omega'}(t_2)  \hat{\rho}_{\ell}  \right] & = 0, \\
\mbox{Tr}_\textrm{line}
\left[  
 \hat{A}_{\omega}^{\dagger}(t_1) \hat{A}_{\omega'}^{\dagger}(t_2)  \hat{\rho}_{\ell}  \right] 
 & =  0.
\end{aligned}
\end{equation}
Then, \cref{eq:BRW-1} reduces to 
\begin{equation}
\frac{d}{dt} \hat{\rho}_I^{(c)}(t)  
= 
G_1 + G_2 + G_3 + G_4 ,
\end{equation}
with
\begin{equation}
\begin{aligned}
G_1 
&=
- 
\int^{t}_{-\infty}\!\!\!\!\! dt'  
\int^{\infty}_0\!\!\!\! d\omega {|\phi(\omega)|}^2
e^{-i\omega\Delta t } 
\hat{C}^{\dagger}_{\omega}(t) 
\hat{C}^{\phantom{g}}_{\omega}(t')
\hat{\rho}_I(t) ,
 \\
G_2
&=
- 
\int^{t}_{-\infty}\!\!\!\!\! dt'  
\int^{\infty}_0\!\!\!\! d\omega {|\phi(\omega)|}^2
e^{i\omega\Delta t } 
\hat{\rho}_I(t) 
\hat{C}^{\dagger}_{\omega}(t') 
\hat{C}^{\phantom{g}}_{\omega}(t),
 \\
G_3
&=
\int^{t}_{-\infty}\!\!\!\!\! dt'  
\int^{\infty}_0\!\!\!\! d\omega {|\phi(\omega)|}^2
 e^{i\omega\Delta t } 
\hat{C}^{\phantom{g}}_{\omega}(t)
\hat{\rho}_I(t) 
\hat{C}^{\dagger}_{\omega}(t') ,
 \\
G_4
&=
\int^{t}_{-\infty}\!\!\!\!\! dt'  
\int^{\infty}_0 \!\!\!\! d\omega {|\phi(\omega)|}^2
 e^{-i\omega\Delta t } 
\hat{C}^{\phantom{g}}_{\omega}(t')
\hat{\rho}_I(t) 
\hat{C}^{\dagger}_{\omega}(t), 
\end{aligned}
\end{equation}
where we introduced $\Delta t = t -t'$.
Using the definition \cref{eq:K_t} for the response 
function $K$, we find
\begin{equation}
\begin{aligned}
G_1 
 & = -
\int^{t}_{-\infty} dt'  
K(\Delta t)
\left[
g_a^2 \hat{n}_a e^{i\omega_a\Delta t} 
+
g_b^2 \hat{n}_b e^{i\omega_b\Delta t} 
\right] \rho_I(t)
\nonumber 
-
\int^{t}_{-\infty} dt'  
K(\Delta t-\tau) g_a g_b e^{i\omega_b t - i\omega_a t'}
\hat{b}^{\dagger} \hat{a} \rho_I(t), \\
G_2 
&=
- 
\int^{t}_{-\infty} dt'  
K^*(\Delta t)
\rho_I(t) \left[
g_a^2 \hat{n}_a e^{-i\omega_a\Delta t} 
+
g_b^2 \hat{n}_b e^{-i\omega_b\Delta t} 
\right] 
-  
\int^{t}_{-\infty} dt'  
K^*(\Delta t-\tau) g_a g_b e^{i\omega_a t' - i\omega_b t}
\rho_I(t) \hat{a}^{\dagger} \hat{b}, \\
 G_3 & =
\int^{t}_{-\infty} dt'  
K^*(\Delta t)
\left[
g_a^2 e^{-i\omega_a\Delta t} 
\hat{a} \rho_I(t) \hat{a}^{\dagger}
+
g_b^2 e^{-i\omega_b\Delta t} 
\hat{b} \rho_I(t) \hat{b}^{\dagger} 
\right]  
+  
\int^{t}_{-\infty} dt'  
K^*(\Delta t-\tau) g_a g_b e^{-i\omega_b t + i\omega_a t'}
\hat{b} \rho_I(t) \hat{a}^{\dagger}, \\
 G_4 &=
\int^{t}_{-\infty} dt'  
K(\Delta t)
\left[
g_a^2 e^{i\omega_a\Delta t} 
\hat{a}  \rho_I(t) \hat{a}^{\dagger}
+
g_b^2 e^{i\omega_b\Delta t} 
\hat{b}  \rho_I(t) \hat{b}^{\dagger}
\right]  
+  
\int^{t}_{-\infty} dt'  
K(\Delta t-\tau) g_a g_b e^{-i\omega_a t' + i\omega_b t}
\hat{a} \rho_I(t) \hat{b}^{\dagger},
\label{eq:G1-G4}
\end{aligned}
\end{equation}
where we used that the integrals containing $K(\Delta t+\tau)$ or $K^*(\Delta t+\tau)$ vanish, since $K(t-t')$ satisfies causality, i.e., $K(t-t') = 0$ for $t'>t$.
Finally, we neglect the frequency and energy dependence of the Fourier transform of the function $K(t-t')$. In the Markovian limit we use \cref{eq:K0delta} and with the definitions of \cref{eq:gagb}, one obtains the master equation
\begin{equation}
\begin{aligned}
 \frac{d}{dt}\rho_I^{(c)}(t)  = &  
-\gamma_a
\left[ \left\{\hat{a}^{\dagger} \hat{a} , \hat{\rho}^{(c)_I} \right\} - 2 \hat{a} \hat{\rho}^{(c)}_I \hat{a}^{\dagger} \right] 
- \gamma_b
\left[ \left\{ \hat{b}^{\dagger} \hat{b} , \hat{\rho}^{(c)_I} \right\} - 2 \hat{b} \hat{\rho}^{(c)}_I \hat{b}^{\dagger} \right] \\
& -2\sqrt{\gamma_a \gamma_b} 
\left[ 
 e^{-i\delta\omega t}
\hat{b}^{\dagger} \hat{a} e^{i\omega_a \tau} \hat{\rho}^{(c)}_I 
+ 
 e^{i\delta\omega t}
\hat{\rho}^{(c)}_I \hat{a}^{\dagger} \hat{b}  
e^{-i\omega_a \tau}
\right]
+ 2\sqrt{\gamma_a \gamma_b} 
\left[ 
 e^{i\delta\omega t}
\hat{b} \hat{\rho}^{(c)}_I \hat{a}^{\dagger } e^{-i\omega_a \tau}
+
 e^{-i\delta\omega t} 
\hat{a} \hat{\rho}^{(c)}_I \hat{b}^{\dagger } e^{i\omega_a \tau}
\right] \,.
\label{eq:Mastereqinte}
\end{aligned}
\end{equation}
using the Ito convention $\int^{t}_{t_0}\!dt' \delta(t-t')= 1/2$ the frequency difference 
$\delta\omega=\omega_a-\omega_b$.
We note that 
the equation for $\hat{\rho}_I^{(c)}(t)$ has 
an explicit time dependence appearing in the mixed terms 
containing both the operators of the two resonators.

Finally, we transform the density matrix from the interaction into to the Schr\"odinger picture
\begin{equation}
\hat{\rho}^{(c)} \left(  t \right) =
e^{- i \hat{H}_\textrm{cas}^{(0)}  t } \hat{\rho}_I^{(c)}\left(  t \right)     e^{ i \hat{H}_\textrm{cas}^{(0)}  t },
\end{equation}
we obtain
\begin{equation}
\begin{aligned}
 \frac{d}{dt} \hat{\rho}^{(c)} \left(  t \right) 
 = & 
- i\left[ \hat{H}_\textrm{cas}^{(0)};  \hat{\rho}^{(c)}(t)  \right]
-\gamma_a
\left[ \left\{\hat{a}^{\dagger} \hat{a} , \hat{\rho}^{(c)} \right\} - 2 \hat{a} \hat{\rho}^{(c)} \hat{a}^{\dagger} \right] 
- \gamma_b
\left[ \left\{ \hat{b}^{\dagger} \hat{b} , \hat{\rho}^{(c)} \right\} - 2 \hat{b} \hat{\rho}^{(c)} \hat{b}^{\dagger} \right] \\
&
-2\sqrt{\gamma_a \gamma_b} 
\left[ 
\hat{b}^{\dagger} \hat{a} e^{i\omega_a \tau} \hat{\rho}^{(c)} 
+ 
\hat{\rho}^{(c)} \hat{a}^{\dagger} \hat{b}  
e^{-i\omega_a \tau}
\right]
+ 2\sqrt{\gamma_a \gamma_b} 
\left[ 
\hat{b} \hat{\rho}^{(c)} \hat{a}^{\dagger } e^{-i\omega_a \tau}
+
\hat{a} \hat{\rho}^{(c)} \hat{b}^{\dagger } e^{i\omega_a \tau}
\right] \,.
\label{eq:L_cascade}
\end{aligned}
\end{equation}
In this way, the explicit time dependence appearing in  
Eq.~(\ref{eq:Mastereqinte})
for $\hat{\rho}_I^{(c)}(t)$ is removed.
In a last step we apply an unitary transformation to redefine the phase of the operator $\hat{a}$
\begin{equation}
\hat{a} \longrightarrow \hat{a}_{\tau} \, = \, \hat{a} \, e^{i\omega_a \tau}.
\end{equation}
In the following, as well as in the main text,  we use 
$\hat{a}$ for the bosonic operator instead of $\hat{a}_{\tau}$.
This corresponds to a shift of the time origin for the evolutions of the two fields which is irrelevant as long one focuses only on the dynamical properties of the mode $b$.
The cascade master \cref{eq:L_cascade}
is equivalent to the one generated by the dissipator $\mathcal{L}_{\textrm{cas}}$ shown in the main text in \cref{eq:cascade} changing the notation 
$\gamma_a \rightarrow \gamma_{\textrm{in}}$ 
and 
$\gamma_b \rightarrow \gamma_{\textrm{c}}$. 
%
%

%%%%%

\section{
Equations for the density matrix elements and analytic formula for 
$\langle n_c(t) \rangle$
} 
\label{app:EoM}

We define the matrix element of the density matrix as 
\begin{equation}
\rho^{c\tilde{c}}_{n_s n_d,\tilde{n}_s \tilde{n}_d} = \left< n_s, n_d,c  \right| \hat{\rho} \left| \tilde{n}_s, \tilde{n}_d, \tilde{c}\right>,
\end{equation}
with $n_s,\tilde{n}_s = 0,1$ and $n_d,\tilde{n}_d = 0,1$ the occupation number for the source and the detection mode in the single photon space, respectively, and $c,\tilde{c}=0,1$ are the QD's occupation states.
The full system's density matrix, including both, the source and detection mode, as well as the QD,  is block-diagonal. 
It splits into a $2\times2$ (zero-photon sector) and a
$4\times4$ block (one-photon sector)
\begin{equation}
    \hat{\rho} 
    \rightarrow
    \left(
    \begin{array}{c c|c c|c c}
        \rho_{00,00}^{00} & \rho_{00,00}^{01} & 0&0&0&0\\
         \rho_{00,00}^{10} & \rho_{00,00}^{11}& 0&0&0&0\\
         \hline \\
          0&0& \rho_{01,01}^{00} & \rho_{01,01}^{01} & \rho_{01,10}^{00} & \rho_{01,10}^{01} \\
           0&0& \rho_{01,01}^{10} & \rho_{01,01}^{11} & \rho_{01,10}^{10} & \rho_{01,10}^{11}\\
           \hline\\
            0&0& \rho_{10,01}^{00} & \rho_{10,01}^{01} & \rho_{10,10}^{00} & \rho_{10,10}^{01}\\
             0&0& \rho_{10,01}^{10} & \rho_{10,01}^{11} & \rho_{10,10}^{00} & \rho_{10,10}^{11}    
    \end{array}
    \right) \, .
\end{equation}
The full equations of motion for each component of the density matrix are generated by \cref{eq:Lindblad}.\\[2mm]
First, we first consider the components diagonal with respect to a photon in the source mode 
$\rho_{10,10}^{00}$ and $\rho_{10,10}^{11}$.
We only report the diagonal elements with respect to charge as the   off-diagonal entries, $\rho_{10,10}^{cc'}$ with $(c\neq c')$, are decoupled from the diagonal ones,  as  for a single QD.
This also holds for the other sub-blocks with different photon occupation.
The corresponding equations read
\begin{equation}
\begin{aligned}
        \dot{\rho}_{10,10}^{00} & = -\left(f(\epsilon) \Gamma_\textrm{QD} + 2\gamma_\textrm{in}\right) \rho_{10,10}^{00}+\left(1 -f(\epsilon) \right)\Gamma_\textrm{QD} \rho_{10,10}^{11},\\
        \dot{\rho}_{10,10}^{11} & = f(\epsilon) \Gamma_\textrm{QD} \rho_{10,10}^{00} - \left[ \left(1 - f(\epsilon) \right)\Gamma_\textrm{QD} + 2\gamma_\textrm{in}\right] \rho_{10,10}^{11}.
    \end{aligned}
\end{equation}
The coupled equations for $\rho_{10,10}^{00}$ and $\rho_{10,10}^{11}$ are closed as expected from the  unidirectional interaction between source and detection mode. 
Using the initial conditions, $\rho_{10,10}^{00}(t=0) = 0, \rho_{10,10}^{11}(t=0) = 1$ 
and $p^{(1)}(t=0)=1$,  
the factorized solutions are
\begin{equation}
\begin{aligned}
    \rho_{10,10}^{00}(t) & = e^{-2\gamma_\textrm{in}t }\left(1 - e^{-\Gamma_\textrm{QD}t} \right)\left(1 - f(\epsilon) \right), \\
    \rho_{10,10}^{11}(t) & = e^{-2\gamma_\textrm{in}t } \left[e^{-\Gamma_\textrm{QD}t}\left(1 - f(\epsilon) \right) + f(\epsilon) \right].
\end{aligned}
\end{equation}
From this we find the average occupation of the source mode
\begin{align}
    \langle \hat{n}_\textrm{in}(t) \rangle = \rho_{10,10}^{00}(t) + \rho_{10,10}^{11}(t) = e^{-2\gamma_\textrm{in}t},
\end{align}
as show in \cref{fig:evolution}.
This solution is independent of the initial choice $\rho_{10,10}^{00}(t=0), \rho_{10,10}^{11}(t=0)$, as long as $\rho_{10,10}^{00}(t~=~0)~+~\rho_{10,10}^{11}(t~=~0)~=~1$ holds.\\[2mm]
Second, we consider the equations for the blocks with  off-diagonal photon states (photon coherence), $ \rho_{10,01}^{00}$ 
and $ \rho_{10,01}^{11}$.
For those, we find
\begin{equation}
    \begin{aligned}
        \dot{\rho}_{10,01}^{00} & = -2\sqrt{\gamma_\textrm{in}\gamma_c} \rho_{10,10}^{00} - \left[\frac{\Gamma_\textrm{QD}}{2}\left(
   f(\epsilon) + f(\epsilon+\lvert \lambda \rvert)
    \right) +  \left(\gamma_\textrm{in} + \gamma_c + i \Delta \omega \right)\right] \rho_{10,01}^{00} + \frac{\Gamma_\textrm{QD}}{2} \left(2 - 
   f(\epsilon) 
    - f(\epsilon+\lvert \lambda \rvert)
    \right) \rho_{10,01}^{11}, \\
    \dot{\rho}_{10,01}^{11} & = -2\sqrt{\gamma_\textrm{in}\gamma_c} \rho_{10,10}^{11}  + \frac{\Gamma_\textrm{QD}}{2}\left(
    f(\epsilon) + 
    f(\epsilon+\lvert \lambda \rvert)
    \right) \rho_{10,01}^{00} - \left[\frac{\Gamma_\textrm{QD}}{2}\left(2 - f(\epsilon) - f(\epsilon+\lvert \lambda \rvert) \right) + \left(\gamma_\textrm{in} + \gamma_c  +i \left(\Delta\omega - \lvert \lambda \rvert\right)\right) \right]\rho_{10,01}^{11}.
    \label{eq:app-offdia-phonon}
        \end{aligned}
\end{equation}
The  equations for $ \rho_{01,10}^{00}$ 
and $ \rho_{01,10}^{11}$ 
are given by the complex conjugate of \cref{eq:app-offdia-phonon}. \\[2mm]
Third, we consider the components diagonal with respect to a having a photon in the detection mode,
\begin{equation}
    \begin{aligned}
        \dot{\rho}_{01,01}^{00} & = - 4 \sqrt{\gamma_\textrm{in}\gamma_c}\, \textrm{Re}\left\{\rho_{10,01}^{00}\right\} -\left(\Gamma_\textrm{QD}
        f(\epsilon+\lvert \lambda \rvert) + 2 \gamma_c\right)\rho_{01,01}^{00}+\Gamma_\textrm{QD}(1-f(\epsilon+\lvert \lambda \rvert))\rho_{01,01}^{11}, \\
        \dot{\rho}_{01,01}^{11} & = - 4 \sqrt{\gamma_\textrm{in}\gamma_c}\, \textrm{Re}\left\{\rho_{10,01}^{11}\right\} +\Gamma_\textrm{QD}f(\epsilon+\lvert \lambda \rvert)  \rho_{01,01}^{00}-\left[\Gamma_\textrm{QD}(1-f(\epsilon+\lvert \lambda \rvert)) + 2 \gamma_c\right]\rho_{01,01}^{11}.
        \label{eq:eom_detectionmode}
    \end{aligned}
\end{equation}
From this, one finds the equation of motion for the occupation of the detection mode
\begin{align}
    \frac{d}{dt}\langle n_c (t) \rangle = \dot{\rho}^{00}_{01,01} + \dot{\rho}^{11}_{01,01}.
    \label{eq:eom_occupation_detection}
\end{align}
This shows the dependence of $\langle n_c \rangle(t)$ on the coherence between source and detection mode (first term in the equation of motion of $\rho^{00}_{01,01}$ and $\rho^{11}_{01,01}$ in \cref{eq:eom_detectionmode}).

Solving the equations of motion of \cref{eq:app-offdia-phonon}, and inserting the solutions into \cref{eq:eom_occupation_detection} leads to an involved expression for  $\langle \hat{n}_c(t)\rangle$. 
We state here the closed form solution for  $T\to0, \Delta \omega = \lvert \lambda \rvert, \gamma_\textrm{in} = \gamma_c$, which is given by
\begin{equation}
    \begin{aligned}
    \langle\hat{n}_c(t)\rangle & = \frac{16\gamma_c^2}{\Gamma_\textrm{QD}^2} e^{-2t\gamma_c}\left(t\Gamma_\textrm{QD} -2 \right) + \frac{16\gamma_c^2}{\lvert \lambda \rvert^2} e^{-2t\gamma_c} \\
        & -\frac{16\gamma_c^2}{\Gamma_\textrm{QD}^2\lvert \lambda \rvert^2} e^{-2t\gamma_c -t \Gamma_\textrm{QD}/2} \cosh\left(\frac{t}{2}\sqrt{\Gamma_\textrm{QD}^2-\lvert \lambda \rvert^2}\right) \left[2\left(\Gamma_\textrm{QD}^2 -\lvert \lambda \rvert^2\right) \cos\left(t\lvert \lambda \rvert/2\right) + 3\lvert \lambda \rvert\Gamma_\textrm{QD}\sin\left(t\lvert \lambda \rvert/2\right)\right] \\
        & + \frac{16\gamma_c^2}{\Gamma_\textrm{QD}^2 \lvert \lambda \rvert^2} e^{-2t\gamma_c -t\Gamma_\textrm{QD}/2}\frac{\sinh\left(\frac{t}{2}\sqrt{\Gamma_\textrm{QD}^2 - \lvert \lambda \rvert^2}\right)}{\sqrt{\Gamma_\textrm{QD}^2 - \lvert \lambda \rvert^2}} \left[\Gamma_\textrm{QD}\left(3\lvert \lambda \rvert^2 - 2\Gamma_\textrm{QD}^2\right)\cos{\left(t\lvert \lambda \rvert/2\right)} + \lvert \lambda \rvert\left(2\lvert \lambda \rvert^2 - 3 \Gamma_\textrm{QD}^2\right)\sin{\left(t\lvert \lambda \rvert/2\right)}\right].
        \label{eq:app-nc_t}
    \end{aligned}
\end{equation}

In the limit $\lvert \lambda \rvert \to 0$ (independent of $T$), i.e, the QD system is decoupled from the photon, 
the general solution of \cref{eq:eom_occupation_detection} yields
\begin{equation}
\begin{aligned}
     \langle \hat{n}_c(t) \rangle^{(\vert \lambda \rvert =0)}
    & =   \frac{8 \gamma_\textrm{in}\gamma_c 
    e^{-2t\left(\gamma_\textrm{in}+\gamma_c\right)} 
    }{\left(\gamma_\textrm{in} - \gamma_c\right)^2 + \Delta \omega^2} \left\{\cosh\left[\left(\gamma_\textrm{in} - \gamma_c\right) t\right] - \cos\left(\Delta\omega t\right)\right\} \\
& =
\begin{cases}
    4 t^2 \gamma_c^2 e^{-2t\gamma_c} & \text{for }  \gamma_\textrm{in} = \gamma_c\, , \Delta\omega = 0 \\
 \frac{16\gamma_c^2}{\Delta \omega^2} \sin^2\left({t \Delta \omega /2}\right)e^{-2\gamma_c t} & \text{for }   \gamma_\textrm{in} = \gamma_c
\, , \Delta\omega \neq 0 \\
\end{cases}
\, . 
    \label{eq:app-nc_t-l0}
\end{aligned}
\end{equation}
For the optimal conditions of maximum
photon occupation of the detection mode, i.e.,  $\gamma_\textrm{in} = \gamma_c$ and $\Delta\omega = 0$, 
the maximum is $\langle n_c \rangle^{(\lvert \lambda \rvert=0)}_\textrm{opt,max} = 4 / e^2$ is at $t = 1/\gamma_c$. 
The  impact of the detuning becomes apparent from the prefactor  $(16\gamma_c^2/\Delta\omega^2)$ 
%, for the case $\gamma_\textrm{in} = \gamma_c$, 
that suppress the detection mode occupation.
A similar prefactor $(16\gamma_c^2/\lvert \lambda \rvert^2)$ appears in Eq.~(\ref{eq:app-nc_t}) for $\langle\hat{n}_c\rangle(t)$ when $\Delta\omega = \lvert \lambda \rvert$, supporting the idea that the dispersive interaction suppresses the occupation by dynamically detuning the detection mode from resonance.
Note that \cref{eq:app-nc_t} to zeroth order in $\lvert \lambda \rvert$ and $\Gamma_\textrm{QD}$ reduces to \cref{eq:app-nc_t-l0} for $(\gamma_\textrm{in} = \gamma_c$ and $\Delta\omega=0$, i.e,
$
    \langle \hat{n}_c(t)\rangle = \langle n_c (t)\rangle^{(\lvert \lambda \rvert=0)}_\textrm{opt} + \mathcal{O}(\lvert \lambda \rvert,\Gamma_\textrm{QD})
$.\\[2mm]
In the limit of $\lvert \lambda \rvert \to \infty$ (and $T \to 0,\  \gamma_\textrm{in} = \gamma_c$) we find for $\Delta \omega=\lvert \lambda \rvert$ that
\begin{align}
    \langle n_c(t) \rangle^{(\lvert \lambda \rvert\to\infty)} = \frac{16\gamma_c^2}{\Gamma_\textrm{QD}^2} e^{-\frac{t}{2}\left(4\gamma_c + \Gamma_\textrm{QD}\right)}\left[ 2 + e^{\frac{t}{2}\Gamma_\textrm{QD}}\left(t\Gamma_\textrm{QD} -2\right)\right],
\end{align}
and its maximum
\begin{align}
   \langle n_c  \rangle^{(\lvert \lambda \rvert\to\infty)}_\textrm{max} = \frac{8\gamma_c}{\Gamma_\textrm{QD} \left(4\gamma_c + \Gamma_\textrm{QD}\right)}e^{-\left(4\gamma_c + \Gamma_\textrm{QD} + 4\gamma_c W(z)\right)/\Gamma_\textrm{QD} }\left[4\gamma_c + \Gamma_\textrm{QD} + 4\gamma_c W(z)\right],
   \label{eq:app-nc_t-linfty}
\end{align}
where $W(z)$ is the Lambert $W$ function at $z= -\exp\left[-1 - \Gamma_\textrm{QD}/4\gamma_c\right] \left(4\gamma_c + \Gamma_\textrm{QD}\right)/4\gamma_c$.\\[2mm]
For completeness, we report the equation for the zero-photon subspace, diagonal in the QD charge
\begin{equation}
    \begin{aligned}
        \dot{\rho}_{00,00}^{00} & = \Gamma_\textrm{QD} \left[\rho_{00,00}^{11} - f(\epsilon)\left(\rho_{00,00}^{00} + \rho_{00,00}^{11}\right)\right] + 2\left[\gamma_\textrm{in} \rho_{10,10}^{00} + 2\sqrt{\gamma_\textrm{in}\gamma_c}\,\textrm{Re}\left\{\rho_{10,01}^{00}\right\} + \gamma_c \rho_{01,01}^{00}\right],\\
        \dot{\rho}_{00,00}^{11} & = \Gamma_\textrm{QD} \left[  f(\epsilon)\left(\rho_{00,00}^{00} + \rho_{00,00}^{11}\right) - \rho_{00,00}^{11}\right] + 2\left[\gamma_\textrm{in} \rho_{10,10}^{11} + 2\sqrt{\gamma_\textrm{in}\gamma_c}\,\textrm{Re}\left\{\rho_{10,01}^{11}\right\} + \gamma_c \rho_{01,01}^{11}\right].
        \label{eq:eom_zerophoton}
    \end{aligned}
\end{equation}
From the solutions for all the components with photons, one can in turn find the solutions for the zero-photon sector in \cref{eq:eom_zerophoton}, and finally for  the occupation of the QD
\begin{equation}
\begin{aligned}
    p^{(0)} &= \rho^{00}_{00,00} + \rho^{00}_{10,10} + \rho^{00}_{01,01},\\
    p^{(1)} & = \rho^{11}_{00,00} + \rho^{11}_{10,10} + \rho^{11}_{01,01}.
\end{aligned}
\end{equation}

\end{widetext}

%%% Bibliography %%%%%%%%%%%%%%%%%%%%%%%%%%%%%%%%%%%%%%%%%%

\bibliography{references}
% add references in file with name references.bib

\end{document}